\documentclass[a4paper,11pt]{article}
\pdfoutput=1 

\usepackage{jheppub} 
\usepackage{graphicx}
\usepackage{hyperref}
\usepackage{cancel}
\usepackage{amssymb}
\usepackage{textcomp}
\usepackage{amsmath}
\usepackage{bm}
\usepackage{times}
\usepackage{epsfig}
\usepackage{color}
\usepackage{ulem}
\usepackage{subfigure}
\usepackage{makecell}

\newcommand{\eV}{{\rm eV}}
\newcommand{\keV}{{\rm keV}}
\newcommand{\MeV}{{\rm MeV}}
\newcommand{\GeV}{{\rm GeV}}
\newcommand{\TeV}{{\rm TeV}}

\newcommand{\pb}{{\rm pb}}

\begin{document}

\title{ Naturally Small Dirac Neutrino Mass with Intermediate $SU(2)_{L}$ Multiplet Fields}

\author[a]{Weijian Wang}
\author[b]{Zhi-Long Han}

\affiliation[a]{Department of Physics, North China Electric Power
University, Baoding 071003,China}
\affiliation[b]{
School of Physics, Nankai University, Tianjin 300071, China}
\date{\today}

\emailAdd{wjnwang96@aliyun.com}
\emailAdd{hanzhilong@mail.nankai.edu.cn}

\abstract{
 If neutrinos are Dirac fermions, certain new physics
beyond the standard model should exist to account for the smallness
of neutrino mass. With two additional scalars and a heavy
intermediate fermion, in this paper, we systematically study the
general mechanism that can naturally generate the tiny Dirac
neutrino mass at tree and in one-loop level. For tree level models,
we focus on natural ones, in which the additional scalars develop
small vacuum expectation values without fine-tuning. For one-loop
level models, we explore those having dark matter candidates under
$Z_2^D$ symmetry. In both cases, we concentrate on $SU(2)_L$
multiplet scalars no larger than quintuplet, and derive the complete
sets of viable models. Phenomenologies, such as lepton flavor
violation, leptogenesis, DM and LHC signatures are briefly discussed.
 }


\maketitle \flushbottom

\section{Introduction}

The mechanism responsible for tiny neutrino mass generation remains
a puzzle. If the neutrinos are Majorana particles, the attractive
scenario is to introduce Weinberg's dimension five operator $\lambda
LL\Phi\Phi/\Lambda$ \cite{wein}, where $\Lambda$ is the typical high
energy scale of underlying new physics. By adding new heavy
intermediate states to the Standard Model (SM) particle content,
there are three canonical mechanisms to realize above operator at
tree level (referred to as type-I, II, III seesaw
models \cite{type1,typ2,typ3}). The smallness of neutrino mass can
also be achieved at low energy scale, either by pushing the mass
operator beyond five
dimension \cite{Babu:2009aq,Picek:2009is,Liao:2010ku,Liao:2010cc,
Ren:2011mh,Kumericki:2011hf,Kumericki:2012bh,Picek:2012ei,McDonald:2013kca,Law:2013gma}
or by attributing the mass term to purely radiative arising at
loop-level (sees Ref.~\cite{1loop,Ma:2006km,2loop,3loop,3loop2} for
classic examples). In these models, new physics may arise at TeV
scale and thus be detectable at LHC or other planned collider
machine \cite{Han:2015hba}. In Ref.~\cite{McDonald:2013kca}, the minimal realizations of
the seesaw mechanisms at tree level are listed according to the
nature of heavy intermediate $SU(2)_{L}$ multiplet fermions. In
Ref.~\cite{Law:2013saa}, the one-loop neutrino mass model proposed by
Ma \cite{Ma:2006km} is generalized to a class of related models with
$SU(2)_{L}$ multiplet fields no larger than adjoint representation.

On the other hand, the experimental evidences establishing whether
neutrinos are of Majorana or Dirac type are still missing. If
neutrinos are Dirac particles and acquire their masses via direct
coupling with SM Higgs boson, the Yukawa coupling constants have to
be unnaturally small in comparison with other SM fermions. To solve
the problem, some mechanism accounting for the smallness of Dirac
neutrino mass have been proposed by many authors at tree(see
Ref.~\cite{Roncadelli:1983ty,Roy:1983be,Chang:1986bp,Mohapatra:1987hh,
Mohapatra:1987nx,Balakrishna:1988bn,Ma:1989tz,Babu:1988yq} for
earlier works and Ref.~\cite{Gu:2006dc,Ma:2014qra,Ma:2015raa,
Valle:2016kyz,Bonilla:2016zef,Chulia:2016ngi,Reig:2016ewy}
for latest works) and loop
level \cite{Gu:2007ug,Farzan:2012sa,Dev:2012sg,Okada:2014vla,
Kanemura:2016ixx,Bonilla:2016diq,Borah:2016zbd,Ma:2016mwh,Ma:2015mjd,Borah:2017leo}.
In Ref.~\cite{Ma:2016mwh}, the generic topographies of diagrams with
specific cases are presented.

In this work, we catalogue the related models that generate the tiny
Dirac neutrino mass at tree and one-loop level. In Sec.~\ref{Tree}, we focus
on the minimal tree level realizations of Dirac seesaws with at most
two extra scalars $S_{1,2}$ and a heavy intermediated Dirac fermion
$F$, see Fig.~\ref{FM:Tree}. As pointed out in Ref.~\cite{Ma:2016mwh}, to
obtain a naturally small Dirac neutrino mass, another symmetry is
required to forbid the
$\overline{\nu}_{L}\nu_{R}\overline{\phi^{0}}$ term, where
$\phi^{0}$ denotes the SM Higgs field. Then the breaking
 of this symmetry induces the effective Dirac neutrino mass
$m_{D}\overline{\nu}_{L}\nu_{R}$. It naively appears that, by adding
appropriate $SU(2)_{L}$ multiplet field variants to SM, there are
infinite ways to realize tree level diagram in Fig.~\ref{FM:Tree}. However,
we will see that, as the Majorana case \cite{McDonald:2013kca}, the
number of candidate models is significantly reduced if only the
models with non-tuning vacuum expectation values (VEVs) are
considered.

At one-loop level, a typical diagram was
proposed \cite{Gu:2007ug,Farzan:2012sa,Ma:2016mwh}, in which the
particle content includes two extra scalars and a gauge-singlet
fermion being odd under $Z_{2}$ symmetry. As a result, the lightest
beyond-SM field is stable and may be considered a dark matter (DM)
candidate. In Sec.~\ref{Loop}, we generalize the approach given in
Ref.~\cite{Law:2013saa}. We list a class of models which generates
the Dirac neutrino masses via the one-loop diagram in Fig.~\ref{FM:Loop} and
simultaneously includes a DM candidate. Without loss of generality,
we mainly restrict our attention on the models with the $SU(2)_{L}$
multiplets fields no larger than adjoint while briefly list the
models with larger multiplets in Appendix. For each model, we
investigate its validity and the type of DM candidate which is
compatible with direct detection experiments. We consider the
phenomenology of the models in Sec.~\ref{PH}. discussing the issues of
lepton number violation processes, leptogenesis and collider
signals. A conclusion is given in Sec.~\ref{CL}.

\section{Tree Level Models for Dirac Neutrino Mass}\label{Tree}

\begin{figure}[!htbp]
\begin{center}
\includegraphics[width=0.5\linewidth]{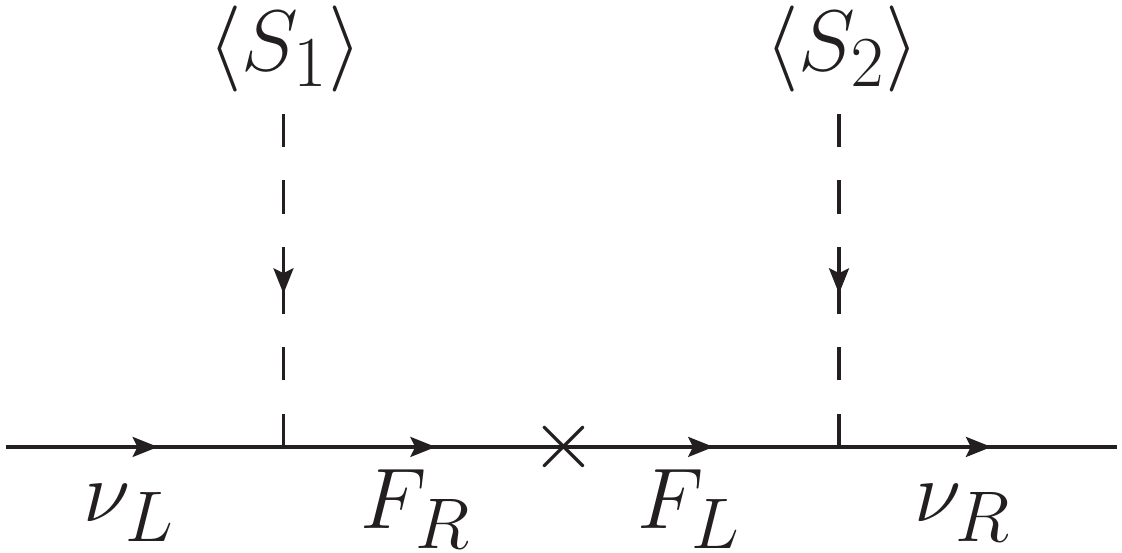}
\end{center}
\caption{Dirac neutrino mass at tree level.} \label{FM:Tree}
\end{figure}

Pathways to naturally small Dirac neutrino mass have been
recently discussed in Ref.~\cite{Ma:2016mwh}. By adding an extra
Dirac fermion singlet/doublet/triplet or scalar doublet, four
tree-level seesaw models are found to realize the Dirac neutrino
mass generation. However, the Dirac seesaw mechanism is more
general when we move beyond the field content given in
Ref.~\cite{Ma:2016mwh}. Following the spirit of Ref.
\cite{McDonald:2013kca,Law:2013gma}, we firstly discuss the
tree-level realization of Dirac seesaw with at most two extra
scalars $S_{1,2}$ and a heavy intermediated Dirac fermion
$F$\footnote{Here, we introduce three generations of
 heavy fermion $F$. For simplicity, we will not show the generation
 indices explicitly in the following discussion.}
(Fig.~\ref{FM:Tree}). Here, the global lepton number symmetry $U(1)_\ell$
is proposed to forbid the unwanted Majorana mass term
$(m_N/2)\overline{\nu_R^C}\nu_R$, meanwhile the discrete $Z_3$
\cite{Ma:2015mjd,Bonilla:2016diq}, $Z_4$
\cite{Heeck:2013rpa,Heeck:2013vha,Chulia:2016ngi} and $\Delta(27)$
\cite{Aranda:2013gga} symmetry are also optional.

\begin{table}[!htbp]
\begin{center}
\begin{tabular}{|c|c|c|c|c|c|c|}
\hline Cases & ~$L_L$~ & ~$S_1$~ & ~$F_R$~ & ~$F_L$~ & ~$S_2$~ &
~$\nu_R$~
\\ \hline
 (A) & $+$ & $-$ & $-$ & $-$ & $+$ & $-$
\\ \hline
(B) & $+$ & $+$ & $+$ & $+$ & $-$ & $-$
\\ \hline
\end{tabular}
\caption{Cases of $Z_2$-charge assignments for relevant fields.}
\label{Tab:Z2}
\end{center}
\end{table}

In order to obtain a naturally small Dirac neutrino mass, another
symmetry $\mathcal{S}$ is required to forbid the
$\bar{\nu}_L\nu_R\overline{\phi^0}$ term. Then the
 broken of this symmetry $\mathcal{S}$ induces the
effective Dirac neutrino mass term $m_D \bar{\nu}_L \nu_R$
\cite{Ma:2016mwh}. The choice of symmetry $\mathcal{S}$ is
model-dependent and here we take the $Z_{2}$ symmetry as an example.
In Table~\ref{Tab:Z2}, we show two possible cases of $Z_2$-charge
assignment for relevant fields. Under the $Z_2$ symmetry, $\nu_R$ is
$Z_2$-odd while other SM particles are $Z_2$-even in all cases,
which is aiming to forbid the $\bar{\nu}_L\nu_R\overline{\phi^0}$
term. Since $F_L$ carries same $Z_2$-charge as $F_R$, $M_F \bar{F}_L
F_R$ is invariant under $Z_2$ as well as SM gauge symmetry.
Therefore, $M_F$ could be assumed to be large. The $Z_2$ symmetry is
broken explicitly by terms as  $H S_1 S_2$, because of opposite
$Z_2$-charge assignment of $S_1$ and $S_2$ for both case (A) and (B)
in Table~\ref{Tab:Z2}.

Some generic features are described from the general tree level
diagram in Fig. \ref{FM:Tree}:
\begin{itemize}
  \item The heavy fermion $F$ is vector-like, which transforms as $F_{L,R}\sim(1,R_F,Y_F)$  under the $SU(3)_C\times SU(2)_L\times U(1)_Y$ gauge symmetry.
  \item The scalars $S_{1,2}$ transform as $S_{1,2}\sim(1,R_{1,2},Y_{1,2})$, and they are necessarily distinct from each other, i.e., $R_1\neq R_2$ or/and $Y_1\neq Y_2$.
  \item The new particles $F$ and $S_{1,2}$ must contain a neutral component, which requires:
      \begin{equation}\label{nt}
      |Y_i|\leq R_i-1,~(i=F,1,2).
      \end{equation}
       And $Y_i$ must be an integer to avoid fractionally charged particles as well.
  \item For isospin allowing to couple $F$ and $S_1(S_2)$ to $L_L(\nu_R)$, following relations should be satisfied:
      \begin{eqnarray} \label{R1}
      R_L\otimes R_1 \supset R_F &~\Rightarrow~& |R_1-R_F|=1,\\ \label{R2}
      R_\nu\otimes R_2 \supset R_F &~\Rightarrow~& ~R_2 = R_F,
      \end{eqnarray}
      where $R_L=2$ and $R_\nu=1$ are the isospin values for SM lepton doublet $L_L$ and neutrino singlet $\nu_R$, respectively.
  \item The neutrality of hyper charge $Y$ then requires that:
     \begin{eqnarray}\label{R3}
      -Y_F+Y_L+Y_1=0 &~\Rightarrow~& Y_1=Y_F+1, \\\label{R4}
      -Y_\nu+Y_F+Y_2=0 &~\Rightarrow~& ~Y_2=-Y_F,
      \end{eqnarray}
      where $Y_L=-1$ and $Y_\nu=0$ are the hyper charges for SM lepton doublet $L_L$ and neutrino singlet $\nu_R$, respectively.
  \item Considering the above relations in Eq.~\ref{R1}--\ref{R4} as well as the fact that the SM Higgs $H$ has the quantum numbers as $R_H=2=R_L$ and $Y_H=1$, one can deduce the following relations:
     \begin{eqnarray}
     (R_H\otimes R_1)\otimes R_2 \supset 1,\\
     Y_1+Y_2-Y_H=0,\label{RH}
     \end{eqnarray}
     which indicates that a trilinear term as $\tilde{H} S_1S_2$ is always allowed in the scalar potential. Here, $\tilde{H}=i\sigma_2 H^{*}$ is the conjugate of the SM Higgs doublet.
\end{itemize}

\begin{figure}[!htbp]
\begin{center}
\includegraphics[width=0.5\linewidth]{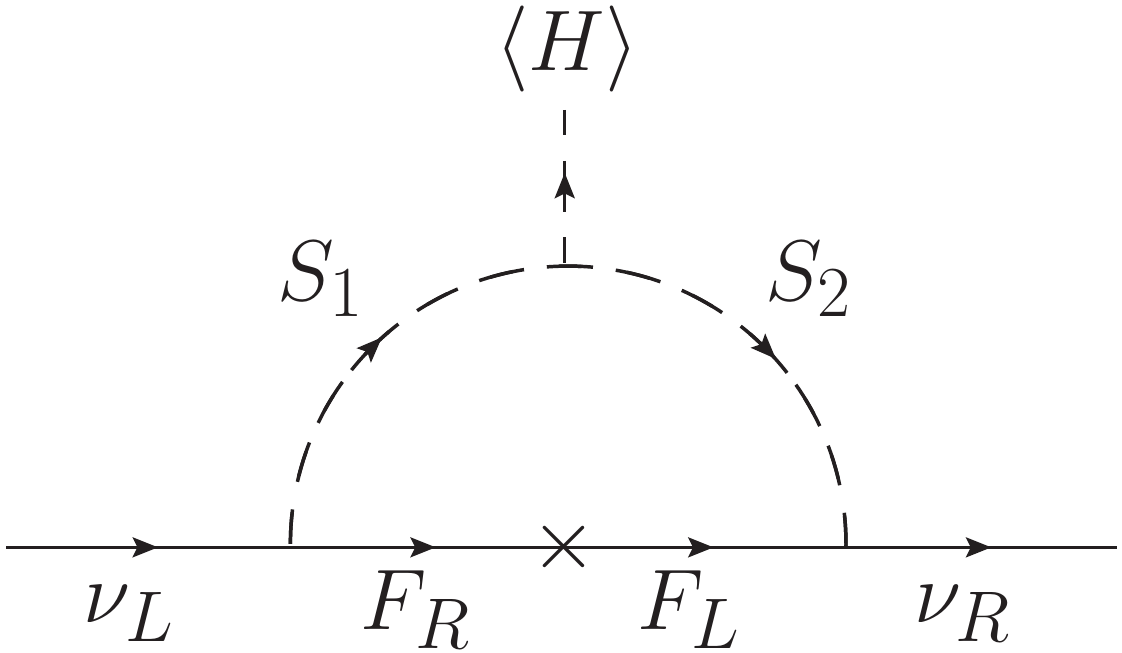}
\end{center}
\caption{Dirac neutrino mass at one-loop level.} \label{FM:Loop}
\end{figure}

We arrive at the relevant terms to generate small Dirac neutrino
masses as shown in Fig.~\ref{FM:Tree}:
\begin{equation}
\mathcal{L} ~\supset~ y_1 \overline{F_R} L_L S_1 + y_2
\overline{\nu_R} F_L S_2 + M_F \overline{F_L} F_R + \mu  \tilde{H}
S_1S_2+\text{h.c.}
\end{equation}
Then the generic form of Dirac seesaw mechanism is realized, for
which the neutrino mass from tree level contribution is given by
\begin{equation}\label{mv-tree}
m_\nu^{\text{tree}} \simeq y_1 y_2 \frac{\langle S_1 \rangle\langle
S_2 \rangle}{M_F},
\end{equation}
For $m_\nu^{\text{tree}}\sim0.1~\eV$, one can set $y_1\sim
y_2\sim10^{-2}$, $\langle S_1\rangle\sim\langle S_2\rangle\sim
10^{-2}~\GeV$, and $M_F\sim 10^2~\GeV$. It is an important issue on
how $S_1$ and $S_2$ develop naturally small VEV comparing to $H$,
which will be discussed in the following. Before proceeding, one
notes that the trilinear $\mu \tilde{H} S_1S_2$ term also
contributes to Dirac neutrino mass via the one-loop diagram in
Fig.~\ref{FM:Loop}. Actually, if the VEVs of $S_1$ and $S_2$ are
forbidden by an additional symmetry, e.g, $Z_2^D$ or $U(1)_D$, only
loop diagram can exist and contribute to the neutrino mass
generation. In this case, it is possible to include dark matter
candidates running in the loop, which is postponed for a more detail
discussion in Sec.~\ref{Loop}.

For the sake of simplicity, one assumes a degenerate mass spectrum
for particles within $F$, $S_1$ and $S_2$, then the one-loop contribution to
Dirac neutrino mass is given by
\begin{equation}\label{mv:loop}
m_{\nu}^{\text{loop}}= C_\nu \frac{\sin2\theta}{32\pi^2} y_1 y_2
M_F\left[\frac{M_{S_2}^2}{M_{S_2}^2-M_F^2}\ln\left(\frac{M_{S_2}^2}{M_F^2}\right)
-\frac{M_{S_1}^2}{M_{S_1}^2-M_F^2}\ln\left(\frac{M_{S_1}^2}{M_F^2}\right)\right],
\end{equation}
where $\theta$ is the mixing angle between $S_1$ and $S_2$. The
coefficient $C_\nu$ is determined by different particle sets running
in the loop and the corresponding Clebsch-Gordan coefficients, thus
is model dependent. For example, in model (a) listed in
Table~\ref{tabnew2} where $F\sim(1,1,0)$, $S_1\sim(1,2,1)$ and
$S_2\sim(1,1,0)$, we have $C_\nu=1$. Depending on relative values
between $M_F$ and $M_{S_{1,2}}$, the expression of
$m_{\nu}^{\text{loop}}$ in Eq.~\ref{mv:loop} can be further
simplified. In the heavy fermion limit with $M_F\gg M_{S_{1,2}}$,
\begin{equation}\label{mv:loop1}
m_{\nu}^{\text{loop}}\simeq C_\nu \frac{\sin2\theta}{32\pi^2}
\frac{y_1 y_2}{M_F}
\left[M_{S_1}^2\ln\left(\frac{M_{S_1}^2}{M_F^2}\right)
-M_{S_2}^2\ln\left(\frac{M_{S_2}^2}{M_F^2}\right)\right].
\end{equation}
While in the opposite limit with $M_F\ll M_{S_{1,2}}$,
\begin{equation}
m_{\nu}^{\text{loop}}\simeq C_\nu \frac{\sin2\theta}{32\pi^2}y_1 y_2
M_F \ln\left(\frac{M_{S_2}^2}{M_{S_1}^2}\right).
\end{equation}
And at last, when $M_F\approx M_{S_{1,2}}$,
\begin{equation}
m_{\nu}^{\text{loop}}\simeq C_\nu \frac{\sin2\theta}{32\pi^2}
\frac{y_1 y_2}{M_F} (M_{S_2}^2-M_{S_1}^2).
\end{equation}
Considering the case of comparable masses with $M_F\approx
M_{S_{1,2}}$ around electroweak scale, we have
\begin{equation}
\frac{m_{\nu}^{\text{loop}}}{m_{\nu}^{\text{tree}}}\sim
\frac{\sin2\theta}{32\pi^2} \frac{M_{S_2}^2-M_{S_1}^2}{\langle S_1
\rangle\langle S_2 \rangle}.
\end{equation}
Therefore, the tree level contribution might be dominant provided
that $\langle S_{1,2}\rangle$ is not too small.

It seems that there could be an infinite number of models satisfying
the above generic features. But when considering constraints from
perturbative unitarity\footnote{ The $2\to2$ tree-level processes of
scalar multiplet pair annihilation into electroweak gauge bosons
receive large contribution for scalar multiplet with large weak
charge. So the upper limits on isospin and hypercharge of a scalar
multiplet can be obtained by requiring that the zeroth partial wave
amplitude satisfies the unitarity bound.}
 \cite{Hally:2012pu,Earl:2013jsa}, we will
concentrate on $SU(2)_L$ scalar multiplet no larger than quintuplet
in this paper. Meanwhile, the number of candidate models could be
significantly reduced when we only consider the models with
non-tuning VEVs for additional scalars $S_{1,2}$
\cite{McDonald:2013kca}. Then, as we shall see, the viable models
are finite and we would like to specify all of them.

First, we consider the simplest case when one of $S_1$ and $S_2$ is
the SM Higgs doublet $H$. Following the conditions in Eq.
\ref{R1}--\ref{R4}, one can figure out four simplest models as:
$S_1=H\sim(1,2,1)$, $F\sim(1,2\mp1,0)$, $S_2\sim(1,2\mp1,0)$ and
$S_1\sim(1,2\mp1,0)$, $F\sim(1,2,-1)$, $S_2=H\sim(1,2,1)$, which
exactly correspond to the cases in Ref.~\cite{Ma:2016mwh}. At the
same time, the trilinear term $\tilde{H} S_1S_2$ becomes $\tilde{H}
H S_{1/2}$ and induces a non-zero VEV of $S_{1/2}$
\begin{equation}
\langle S_{1/2} \rangle \simeq \mu\frac{\langle H
\rangle^2}{M^2_{S_{1/2}}}.
\end{equation}
Notably, the trilinear term $\tilde{H}H S_{1/2}$ could be an
explicit $Z_2$ breaking term as in the cases (A) and (B) shown in
Table~\ref{Tab:Z2}. Thus the small $\langle S_{1/2} \rangle$ is
acquired in the technically natural limit of $\mu\ll\langle H
\rangle$ even for $M_{S_{1/2}}$ around electroweak scale. Then the
tree level neutrino mass in Eq. \ref{mv-tree} is expressed as:
\begin{equation}
m_\nu^{\text{tree}}\simeq y_1y_2 \frac{\mu\, \langle H\rangle^3}{M_F
M_{S_{1/2}}^2}.
\end{equation}
Typically, we can acquire $m_\nu^{\text{tree}}\sim0.1~\eV$ by seting
$y_{1,2}\sim10^{-2}$, $\mu\sim1~\MeV$, and $M_F\sim M_{S_{1/2}}\sim
1~\TeV$. Provided $M_{S_{1/2}}\sim M_{F}=M$, the tiny tree level
Dirac neutrino mass is generated from a dimension $d=6$ effective
low-energy operator as $\mathcal{O}_\nu= \mu \overline{\nu_R} L_L
H^3/M^3$.

Notably, a special case, the so-called neutrinophilic two Higgs
doublet model ($\nu$2HDM), appears in
literature\cite{Ma:2000cc,Davidson:2009ha,Bonilla:2016zef}, where
the new scalar doublet $\eta\sim(1,2,1)$ transforms the same as SM Higgs
doublet under SM gauge group but carries some new charge, i.e.,
$Z_2$ or $U(1)$~\cite{Haba:2011nb,Machado:2015sha,Wang:2016vfj}. In
this model, the new Yukawa coupling $\overline{\nu_R}
\tilde{\eta}^\dag L_L$ is allowed and the small VEV of $\eta$ can be
obtained by adding a soft $Z_2$ or $U(1)$ breaking term as
$\eta^\dag H$, leading to naturally small Dirac neutrino masses
\cite{Ma:2016mwh}. In this case, the tiny Dirac neutrino mass is
generated from a dimension $d=4$ effective low-energy operator as
$\mathcal{O}_\nu = \mu\overline{\nu_R} \tilde{H}^\dag  L_L/M$.

Now we move beyond the simplest case and explore further generations
with both $S_1$ and $S_2$ being new particles. After the electroweak
symmetry breaking, VEVs of $S_{1,2}$ will usually contribute to the
$W$ and $Z$ boson masses. Especially, for those scalars with
$SU(2)_{L}$ representation $R_{1,2}>2$, their VEVs $\langle
S_{1,2}\rangle$ will affect the $\rho$ parameter away from the SM
value $\rho=1$ at tree-level, which then leads to tight bound on
$\langle S_{1,2}\rangle \lesssim\mathcal{O}(1)~\GeV$
\cite{Olive:2016xmw}. The trilinear $\tilde{H}S_1S_2$ term alone can
not ensure that both $\langle S_{1,2} \rangle$ are naturally small
in general case. Therefore, in order to produce non-tuning VEVs for
$S_{1,2}$, the scalar potential $V(H,S_1,S_2)$ should contain linear
$S_1$ or/and $S_2$ terms as $S_{1,2}H^n(n\leq3)$. With these
conditions in mind, we find the $S_1$ or/and $S_2$ with the quantum
numbers as(see Ref.~\cite{McDonald:2013kca} for more details):
\begin{equation}\label{qn}
S_{1,2}\sim (1,2,\pm1),(1,3,0), (1,3,\pm2), (1,4,\pm1),(1,4,\pm3).
\end{equation}
On the other hand, if only $S_i$ obtains a naturally small VEV from
the term $S_{i}H^n(n\leq3)$, the trilinear $\tilde{H}S_1S_2$ term
will induce a naturally suppressed VEV for $S_j$ as:
\begin{equation}\label{vj}
\langle S_j \rangle \simeq \mu \frac{\langle S_i\rangle \langle H
\rangle}{M_{S_j}^2},~\text{for}~i\neq j.
\end{equation}
Thus we expect $\langle S_j \rangle\lesssim \langle S_i \rangle$,
when $\mu\lesssim\langle H \rangle\lesssim M_{S_j}$.

Based on the above statement, the general strategy for determining a
specific Dirac seesaw model is quite straight. First, one determines
a scalar $S_i$ with quantum number in Eq.~\ref{qn}, and then the
viable sets of quantum numbers for $F$ and $S_j$ can be obtained by
Eq.~\ref{R1}--\ref{R4}. Following this procedure, we have listed all
viable models in Table~\ref{tabnew1}. Clearly from Table
\ref{tabnew1}, naturally small Dirac neutrino mass arises from even
number dimension effective operators as $\mathcal{O}_\nu =
\overline{\nu_R} L_L H^{2n+1}/\Lambda^{2n}$ and a higher
scalar/fermion representation generally tends to a higher dimension
effective operator.

\begin{table}[!htbp]
\begin{center}
\begin{tabular}{|c|c|c|c|c|c|}\hline\hline
Models & $F$ & $S_{1}$ & $S_{2}$ & $[\mathcal{O}_\nu]$ & style\\
\hline (A) & $(1,1,0)$ & $H(1,2,1)$ & $\phi(1,1,0)$ & $d=6$ & minimal\\
\hline (B) & $(1,2,-1)$ & $\phi(1,1,0)$ & $H(1,2,1)$ & $d=6$ & minimal \\
\hline (C) & $(1,2,-1)$ & $\Delta(1,3,0)$ & $H(1,2,1)$ & $d=6$ & minimal \\
\hline (D) & $(1,2,1)$ & $\Delta(1,3,2)$ & $\eta(1,2,-1)$ & $d=(4)6$ & (non-)minimal \\
\hline (E) & $(1,3,0)$ & $H(1,2,1)$ & $\Delta(1,3,0)$ & $d=6$ & minimal\\
\hline (F)& $(1,3,0)$ & $\chi(1,4,1)$ & $\Delta(1,3,0)$ & $d=6,8$&
non-minimal\\
\hline (G)& $(1,3,-2)$ & $\eta(1,2,-1)$ & $\Delta(1,3,2)$ & $d=(4)6$ &(non-)minimal\\
\hline (H)& $(1,3,-2)$ & $\chi(1,4,-1)$ & $\Delta(1,3,2)$  & $d=(6)8$&(non-)minimal\\
\hline (I)& $(1,3,2)$ & $\chi(1,4,3)$ & $\Delta(1,3,-2)$ & $d=8$ &
minimal\\
\hline (J)& $(1,4,1)$ & $\Delta(1,3,2)$ & $\chi(1,4,-1)$ & $d=8$&
minimal\\
\hline (K)& $(1,4,1)$ & $\Phi(1,5,2)$ & $\chi(1,4,-1)$ & $d=10$&
minimal\\
\hline (L)& $(1,4,-1)$ & $\Delta(1,3,0)$ & $\chi(1,4,1)$ & $d=8$&
minimal\\
\hline (M)& $(1,4,-1)$ & $\Phi(1,5,0)$ & $\chi(1,4,1)$ &$d=10$&
minimal\\
\hline (N)& $(1,4,3)$ & $\Phi(1,5,4)$ & $\chi(1,4,-3)$ & $d=10$&
minimal\\
\hline (O)& $(1,4,-3)$ & $\Delta(1,3,-2)$ & $\chi(1,4,3)$ & $d=10$&
minimal\\
\hline (P)& $(1,4,-3)$ & $\Phi(1,5,-2)$ & $\chi(1,4,3)$ & $d=10$&
minimal\\
\hline (Q)& $(1,5,0)$ & $\chi(1,4,1)$ & $\Phi(1,5,0)$ & $d=10$&
minimal\\
\hline (R)& $(1,5,2)$ & $\chi(1,4,3)$ & $\Phi(1,5,-2)$ & $d=10$&
minimal\\
\hline (S)& $(1,5,-2)$ & $\chi(1,4,-1)$ & $\Phi(1,5,2)$ & $d=10$ &
minimal\\
\hline (T)& $(1,5,-4)$ & $\chi(1,4,-3)$ & $\Phi(1,5,4)$ & $d=10$&
minimal\\
\hline\hline
\end{tabular}
\caption{Natural tree level seesaws for Dirac neutrinos. For
simplicity, we denote new scalar singlet to quintuplet as $\phi$,
$\eta$, $\Delta$, $\chi$ and $\Phi$, respectively.} \label{tabnew1}
\end{center}
\end{table}

Some comments on specific models are as following. Model (A) and (B)
contains a scalar singlet $\phi\sim(1,1,0)$. In our consideration,
VEV of $\phi$ is induced from the $Z_2$ breaking trilinear term $\mu
\tilde{H}H\phi$, thus $\langle \phi\rangle\simeq \mu \langle
H\rangle^2/M_{\phi}^2$ is naturally small when $\mu\ll\langle
H\rangle\lesssim M_{\phi}$. Since the VEV of $\phi$ does not
contribute to the $\rho$-parameters, it may be typically around
electro-weak scale and originated from the spontaneous breaking of
scalar potential \cite{Chulia:2016ngi,Robens:2015gla}. In this way, one needs the
intermediate fermion $F$ with $\mathcal{O}(10^{10})~\GeV$ mass scale
to generate proper neutrino masses, just as the canonical type-I
seesaw model.

Model (D) and (G) employ a scalar doublet $\eta\sim(1,2,-1)$.
Provided $\eta$ is $Z_2$-odd as $\nu_R$ (as case (B)), the Yukawa
coupling $\overline{\nu_R}\eta^\dag L_L $ is allowed. After $\eta$
develops a VEV from the soft term $\tilde{\eta}^\dag H$, the
$\overline{\nu_R}\eta^\dag L_L $ term induces a Dirac mass term
corresponding to dimension $d=4$ effective operator as
$\mathcal{O}_\nu = \overline{\nu_R} \tilde{H}^\dag L_L$. Meanwhile,
the heavy intermediate fermion $F$ together with $\eta$ and another
scalar triplet $\Delta~(1,3,2)$ generate Dirac neutrino mass from
$d=6$ effective operator as $\mathcal{O}_\nu = \overline{\nu_R} L_L
H^3/\Lambda^2$. Thus, light Dirac neutrino mass have two
contributions, which can be written as
\begin{equation}\label{DG}
m_\nu^\text{tree}\simeq y_{1/2} \langle\eta\rangle + y_1
y_2\frac{\langle\eta\rangle\langle\Delta\rangle}{M_F}.
\end{equation}
Since $\langle\Delta\rangle\ll M_F$, the Dirac neutrino mass is
dominant by the  first term in Eq.~\ref{DG}. Hence, model (D) and
(G) are non-minimal, and can be regarded as just a more complicated
extension of the $\nu$2HDM with new contributions to Dirac neutrino
mass subdominant.

In contrast, if $\eta$ is $Z_2$-even as case (A) shown in
Table~\ref{Tab:Z2}, we might be able to treat $\eta$ as the
charge-conjugate field of $H$, i.e., $\eta=\tilde{H}$. In this case,
the Yukawa coupling $\overline{\nu_R}\eta^\dag L_L=\overline{\nu_R}
\tilde{H}^\dag L_L $ is forbidden, so the light Dirac neutrino mass
can only be induced by the heavy intermediate fermion $F$ as
\begin{equation}
m_\nu^\text{tree}\simeq y_1 y_2\frac{\langle
H\rangle\langle\Delta\rangle}{M_F}.
\end{equation}
Then in addition to the four obvious minimal models--model (A), (B),
(C) and (E), we get two more minimal models--model (D) and (G) with
$d=6$ effective operators as well. When counting on the heavy
intermediate fermion, one more representation $F\sim(1,3,-2)$ is
employed in model (G), while we can regard $F\sim(1,2,1)$ in model
(D) as the charge-conjugate of $F\sim(1,2,-1)$ in model (B) and (C).

For the scalar triplet $\Delta\sim(1,3,\pm2)$ which is involved in
model (D), (G), (H), (I) and (J), the $L_L L_L \Delta$ term is
forbidden by the unbroken $U(1)_\ell$ lepton symmetry in case of
Dirac neutrino. Since for tree level models in this work, we can
only assign lepton number $L=1$ or $L=0$ to new fermion $F$ and
scalars $S_{1,2}$ (including $\Delta$), respectively.

Comparing with model (E) and (F), it is obvious that model (F) is
essentially  model (E) with an additional scalar quadruplet
$\chi\sim(1,4,1)$. As a result, model (F) is non-minimal, and
neutrino mass is generated by two distinct tree level diagrams. And
provided $\eta=\tilde{H}$ in model (G), then model (H) is clearly
also non-minimal. Under such circumstances, the light neutrino mass
for model (F) and (H) is given by:
\begin{equation}
m_\nu^\text{tree}\simeq y^H_1 y_2 \frac{\langle H \rangle \langle
\Delta \rangle}{M_F} + y^{\chi}_1 y_2 \frac{\langle \chi \rangle
\langle \Delta \rangle}{M_F},
\end{equation}
which correspond to effective operators of dimension $d=6$ and
$d=8$, respectively.

Now let's look at a specific model, i.e., model (K) in Table
\ref{tabnew1}, to show how to construct a complete model. First, we
choose $S_2\sim(1,4,-1)$ in Eq.~\ref{qn}. Then, the quantum number
of $F\sim(1,4,1)$ can be obtained from constraints in Eq.~\ref{R2},
\ref{R4} by the Yukawa coupling $\overline{\nu_R} F_L S_2$. At last,
inspection of constraints in Eq.~\ref{R1}, \ref{R3} by the other
Yukawa coupling $\overline{F_R}L_L S_1$ reveals that either
$S_1\sim(1,3,2)$ or $S_1\sim(1,5,2)$ corresponding to model (J) and
(K), respectively.

\begin{figure}[!htbp]
\begin{center}
\includegraphics[width=0.5\linewidth]{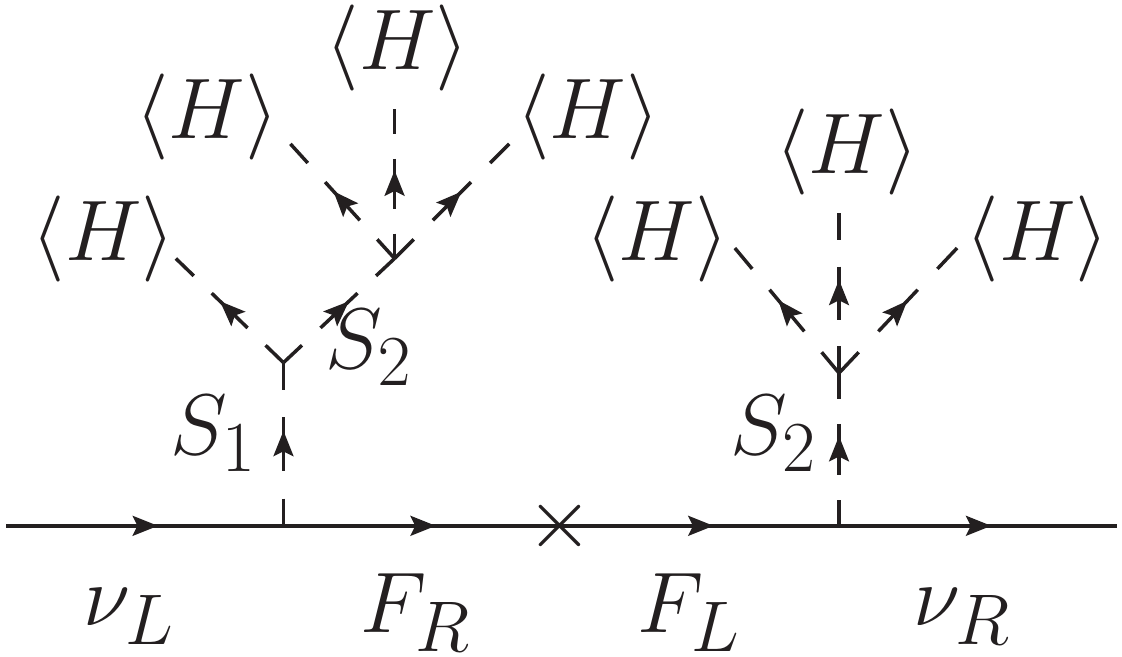}
\end{center}
\caption{Tree level Dirac seesaw of model (K) in Table
\ref{tabnew1}.} \label{FM:MK}
\end{figure}

In Fig.~\ref{FM:MK}, we depict the tree level Dirac seesaw of model
(K). The scalar quadruplet $S_2\sim(1,4,1)$ acquires a naturally
small VEV from the quartic term $\lambda S_{2}^\dag HH^\dag
\tilde{H}$ as:
\begin{equation}
\langle S_2 \rangle \simeq \lambda \frac{\langle H
\rangle^3}{M_{S_2}^2}.
\end{equation}
For the scalar quintuplet $S_1\sim (1,5,2)$, the trilinear
$\tilde{H}S_1S_2$ term ensures $S_1$ also develops a naturally small
VEV as shown in Eq.~\ref{vj}:
\begin{equation}
\langle S_1 \rangle \simeq \mu \frac{\langle S_2\rangle \langle H
\rangle}{M_{S_1}^2} \simeq \lambda \frac{\mu\, \langle H
\rangle^4}{M_{S_1}^2 M_{S_2}^2}.
\end{equation}
As a consequence, the tree level Dirac neutrino mass in model K is:
\begin{equation}
m_\nu^{\text{tree}}\simeq y_1y_2 \frac{\langle S_1 \rangle\langle
S_2 \rangle}{M_F} \simeq y_1y_2 \frac{\lambda \mu}{M_F}
\frac{\langle H \rangle^7}{M_{S_1}^2 M_{S_2}^4}.
\end{equation}
Supposing $\mu\sim M_{S_{1,2}}\sim M_F =M$, then we have
$m_{\nu}^{\text{tree}}\propto \langle H\rangle^7/M^6$, which
indicates that the tiny Dirac neutrino mass is induced by a
dimension $d=10$ effective operator as $\mathcal{O}_\nu =
\overline{\nu_R} L_L H^7/M^6$.

\section{One-loop Models for Dirac Neutrino Mass}\label{Loop}

Now, we move forward to the purely radiative generation of Dirac
neutrino mass. There were variant of models proposed in this direction.
At one-loop level, the simplest model discussed in
Ref.\cite{Kanemura:2011jj} is based on soft-broken $Z_{2}$ symmetry,
where two new charged scalar singlets are employed. However, no DM
candidate can be incorporated in this model. Another appealing way
is introducing an additional symmetry, e.g, a dark $Z_{2}$
symmetry($Z_{2}^{D}$), under which $S_{1,2}$ and $F$ carry
$Z_2^D$-odd charge while all SM fields transform trivially. In this
way, the VEVs of $S_{1,2}$ is forbidden and neutrino mass can only
be generated via the one-loop diagram shown in Fig.~\ref{FM:Loop}. Due to
the $Z_{2}^{D}$ odd protection, the lightest neutral component
within the inert fields $S_{1,2}$ or $F$ is stable, and thus becomes
a DM candidate. In this paper, we restrict our attention on the
models involving neutral components and analyze their validity as a
DM. We focus on models with representations no larger than the
adjoint representation, and then briefly discuss larger multiplets
with quadruplet and/or quintuplet of $SU(2)_L$.

From Fig.~\ref{FM:Loop}, it reveals that, with slightly modification
of statements around Eq.~\ref{nt}, they are still applicable for one-loop
case. The difference comes from the fact that in loop models the
neutral field does not have to propagate inside the loop, hence we
only require that at least one of the new fields $S_{1,2}$ and $F$
has a neutral content. The other constraints, i.e.,
Eq.~\ref{R1}--\ref{RH} , are directly coming or indirectly derived
from the relevant Yukawa coupling, so they are still capable for
loop models.

With the comments given above, a systematic analysis is made to
exhaust the models that generate Dirac neutrino mass via
Fig.~\ref{FM:Loop}. In Table~\ref{tabnew2}, we list all possible
distinct models with representations no larger than the adjoint
representation. There are totally ten viable models, and the
simplest three of them, i.e., model (a), (c) and (e), are already
mentioned in Ref.~\cite{Ma:2016mwh}. For models with quadruplet
and/or quintuplet, we depict them in Table~\ref{tabnew3}.

\begin{table}[!htbp]
\begin{center}
\begin{tabular}{|c|c|c|c|c|}\hline\hline
Models & $F$ & $S_{1}$ & $S_{2}$ & $Z_2^D$ DM \\
\hline (a) & $(1,1,0)$ & $\eta(1,2,1)$ & $\phi(1,1,0)$ & Inert Singlet or Doublet \\
\hline (b) & $(1,1,-2)$ & $\eta(1,2,-1)$ & $\phi(1,1,2)$ & Inert Doublet  \\
\hline (c) & $(1,2,-1)$ & $\phi(1,1,0)$ & $\eta(1,2,1)$ & Inert Singlet or Doublet  \\
\hline (d) & $(1,2,-1)$ & $\Delta(1,3,0)$ & $\eta(1,2,1)$ & Inert Doublet or Triplet  \\
\hline (e) & $(1,2,1)$ & $\phi(1,1,2)$ & $\eta(1,2,-1)$ & Inert Doublet \\
\hline (f) & $(1,2,1)$ & $\Delta(1,3,2)$ & $\eta(1,2,-1)$ & Inert Doublet or Triplet \\
\hline (g) & $(1,2,-3)$ & $\Delta(1,3,-2)$ & $\eta(1,2,3)$ & Excluded \\
\hline (h) & $(1,3,0)$ & $\eta(1,2,1)$ & $\Delta(1,3,0)$ & Inert Doublet or Triplet \\
\hline (i) & $(1,3,-2)$ & $\eta(1,2,-1)$ & $\Delta(1,3,2)$ & Inert Doublet or Triplet  \\
\hline (j) & $(1,3,2)$ & $\eta(1,2,3)$ & $\Delta(1,3,-2)$ & Excluded  \\
\hline\hline
\end{tabular}
\caption{Radiative neutrino mass for Dirac neutrinos with DM
candidate.} \label{tabnew2}
\end{center}
\end{table}

The generic one-loop Dirac neutrino mass matrix has already been
given in Eq.~\ref{mv:loop}. The trilinear term $\mu \tilde{H}S_1S_2$
still induces the mixing between inert scalars $S_1$ and $S_2$,
meanwhile the newly employed $Z_2^D$ symmetry forbids the mixing
between $S_{1,2}$ and SM Higgs doublet $H$. To acquire
$m_\nu^\text{loop}\sim0.1\eV$, we can set $y_1\sim
y_2\sim\theta\sim10^{-3}$ with all inert particles around
$\mathcal{O}(\TeV)$.

A detailed study on the DM phenomenology for all models presented in
Table~\ref{tabnew2} and \ref{tabnew3} is beyond the scope of this
paper. First, we briefly discuss viable DM candidate in specific
models in this section. Then in the next section, we choose model
(a) as our benchmark model for a more detail study.

First, we turn our intention into fermion DM candidate. In model
(a), an inert fermion singlet $F\sim(1,1,0)$ is introduced. Possible
annihilation channels are: 1), $F\bar{F}\to
\ell^+\ell-,\nu\bar{\nu}$ mediated by $\eta$ via the Yukawa coupling
$y_1$; 2), $F\bar{F}\to \nu\bar{\nu}$ mediated by $\phi$ via the
Yukawa coupling $y_2$; 3), coannihilation with $\eta(\phi)$, when
$M_{\eta(\phi)}$ is close to $M_{F}$. For all these channels, not
too small
 Yukawa couplings $y_1$ and/or $y_2$ of $\mathcal{O}(0.1)$ are required to
generate the correct relic density
\cite{Kubo:2006yx,Vicente:2014wga}.
 On the other hand, for $\eta$
involved channels, the Yukawa coupling $y_1$ receives tight
constraints from lepton flavor violating processes
\cite{Vicente:2014wga}. Therefore, we expect that the Yukawa
couplings satisfy the relation $y_1\ll y_2$, which further indicates
that the $\phi$ mediated process is dominant provided
$M_{\phi}\approx M_{\eta}$.

Another notable model with viable fermion DM is model (h), where an
inert fermion triplet $F\sim(1,3,0)$ is introduced. The neutral
component $F^0$ can serve as a DM candidate. Due to its electroweak
couplings to gauge bosons, the relic density of $F^0$ is dominantly
determined by the annihilation and co-annihilation of itself and
$F^\pm$, which requires that $M_{F^0}$ is around $2.6~\TeV$
\cite{Ma:2008cu,Chao:2012sz,vonderPahlen:2016cbw}. In this case,
$S_1\sim(1,2,1)$ and $S_2\sim(1,3,0)$ should be heavier than
$2.6~\TeV$, thus hardly being tested at LHC.

Fermion DM in other models with no larger than adjoint
representation are excluded. Clearly, for model (b) and (g),
$F\sim(1,1,-2)$ and $F\sim(1,2,-3)$ do not have neutral component,
thus these two models do not have fermion DM candidate. On the other
hand, model (c), (d), (e), (f) employ $F\sim(1,2,\pm1)$ and model
(i), (j) employ $F\sim(1,3,\pm2)$, which contain neutral fermions.
But all the neutral fermions in these model have non-zero
hypercharge, which will lead to detectable DM-nucleon scattering
cross section via $Z$-boson exchange. So they have already been
excluded by direct detection experiments, such as, LUX
\cite{Akerib:2013tjd} and PandaX-II \cite{Tan:2016zwf}.

Then we move onto scalar dark matter. Considering the constraints
from LFV and tiny neutrino masses, it is better to set the Yukawa
coupling $y_1\sim y_2\lesssim10^{-2}$. In this way, the contribution
of heavy fermion $F$ to scalar DM variables is negligible. Both
model (a) and (c) introduce an inert scalar singlet
$\phi\sim(1,1,0)$ \cite{Cline:2013gha} and an inert scalar doublet
$\eta\sim(1,2,1)$ \cite{Arhrib:2013ela}. In principle, either of
$\phi$ and $\eta$ can solely paly the role of dark matter candidate
under the $Z_2^D$ symmetry. In these two models, the trilinear term
$\mu \phi \eta^\dag H/\sqrt{2}$ will induce the mixing between
$\phi$ and $\eta_R^0$, and the allowed parameter space thus are
expected enlarged. Detail phenomenological aspects for inert
singlet-doublet scalar dark matter can be found in
Ref.~\cite{Cohen:2011ec,Kakizaki:2016dza}. From the result of
Ref.~\cite{Cohen:2011ec,Kakizaki:2016dza}, we know that the mixing
angle $\theta$ between $\phi$ and $\eta_R^0$ must be small enough to
avoid too large DM-nucleon scattering cross section if DM is
dominant by $\phi$ component. Notably, there exists a value of
$\sin\theta$ for which the correct relic density is maintained only
via the four-point gauge interactions when $M_{\phi}>M_W$. Around
this point, the spin-independent detection cross section drops
dramatically, since the only tree level contribution from the Higgs
boson vanishes. Meanwhile, for $M_{\phi}<M_W$, the relic density is
determined by the Higgs portal. And current direct detection
experiments requires that $M_{\phi}\approx M_h/2$ should be
satisfied for light DM \cite{He:2016mls}. On the other hand if the
dark matter is dominant by $\eta$ component, either $\eta_R^0$ or
$\eta_I^0$, then a mass splitting $\Delta
M=|M_{\eta_R^0}-M_{\eta_I^0}|>100~\keV$ between $\eta_R^0$ and
$\eta_I^0$ is required to escape the direct detection bound. In
these two models, the required mass splitting can be obtained by
choosing curtain values of $\mu$ in the trilinear term $\mu \phi
\eta^\dag H$ and $\kappa$ in the quartic term $\kappa (\eta^\dag
H)^2$ \cite{Gu:2007ug}.

For model (b) and (e), the only DM candidate comes from the inert
scalar doublet $\eta\sim(1,2,-1)$ \cite{Arhrib:2013ela}, since the
other inert scalar $\phi\sim(1,1,2)$ is a charged scalar singlet. It
is noted that in both models, the neutral components $\eta_R^0$ or
$\eta_I^0$ does not contribute to radiative neutrino mass.
Considering the fact that small mixing angle $\theta$ between
$\phi^\pm$ and $\eta^\pm$ are favored by neutrino mass, the DM
phenomenology of $\eta$ will be quite similar as a standard inert
doublet model. Under constraints from relic density, direct
detection and indirect detection, there are two mass region allowed
for $M_{\eta_R^0/\eta_I^0}$: one is the low mass region with
$50~\GeV \lesssim M_{\eta_R^0/\eta_I^0}\lesssim70~\GeV$, and the
other is the high mass region with $500~\GeV \lesssim
M_{\eta_R^0/\eta_I^0}$ \cite{Arhrib:2013ela}. For the light mass
region, pair and associated production processes as $\eta^+\eta^-$
and $\eta^\pm \eta^0_R/\eta^0_I$ will lead to multi-lepton plus
missing transverse energy $\cancel{E}_T$ signatures at LHC, which
has been extensively studied in Ref.~\cite{Dolle:2009ft}. While for
the high mass region, although hard to be test at LHC, most
parameter space of this region is in the reach of CTA experiment
\cite{Queiroz:2015utg}.

For model (d) and (h), they employ an inert doublet
$\eta\sim(1,2,1)$ and a real inert scalar triplet
$\Delta\sim(1,3,0)$\cite{Cirelli:2005uq,FileviezPerez:2008bj,Araki:2011hm}.
Alternatively, the DM candidate could be either $\eta_R^0/\eta_I^0$
in the inert doublet $\eta$ or $\Delta^0$ in the inert triplet
$\Delta$ \cite{Lu:2016ucn}. If the mass of inert triplet $\Delta$ is
much heavier than the inert doublet $\eta$, we again arrive at the
well studied inert doublet model \cite{Arhrib:2013ela} as just
discussed above. Here, we consider the opposite case where
$\Delta^{0}$ is lighter than the inert doublet $\eta_R^0/\eta_I^0$,
and serve as the DM candidate. Determined by the DM relic density,
$M_{\Delta^0}$ is found to be around $2.5~\TeV$ if (co-)annihilation
is via pure gauge coupling, meanwhile the scalar interactions could
push $M_{\Delta^0}$ up to about $20~\TeV$ due to the Sommerfeld
effect\cite{Cirelli:2005uq}. Since $\Delta^0$ does not interact with
$Z$-boson, the DM-nucleon scattering process through the exchange of
SM Higgs $h$ at tree level and gauge bosons at one-loop level.
 And the spin-independent DM-nucleon scattering cross section
at one-loop level is calculated as \cite{Lu:2016ucn}
\begin{equation}
\sigma_{\text{SI}}=\frac{g_2^8}{256\pi^3}\frac{f_N^2 M_N^4}{M_W^2}
\left[\frac{R_\Delta^2-1}{8}\left(\frac{1}{M_W^2}+\frac{1}{M_h^2}\right)
-\frac{16\pi}{g_2^4}\frac{\lambda_{h\Delta^0}}{M_h^2}\frac{M_W}{M_{\Delta^0}}
\right]^2
\end{equation}
Here, $g_2$ is the $SU(2)_L$ gauge coupling, $f_N=0.3$ is the nucleon matrix element,
$M_N=939~\MeV$ is the average nucleon mass, $R_\Delta=3$ is the dimension of
inert triplet $\Delta$, and $\lambda_{h\Delta^0}$ is the coupling between
the DM $\Delta^0$ and SM Higgs $h$. For vanishing DM-Higgs coupling $\lambda_{h\Delta^0}=0$,
the spin-independent cross section $\sigma_{\text{SI}}$ is
$9\times10^{-10}~\pb$, which is lower than current LUX bound \cite{Akerib:2013tjd}.
Remarkably, for certain DM-Higgs coupling, i.e.,
\begin{equation}
\lambda_{h\Delta^0}=\frac{R_\Delta^2-1}{8}\frac{g_2^4 M_{\Delta^0}}{16\pi M_W}
\left(1+\frac{M_h^2}{M_W^2}\right)\approx 0.4
\end{equation}
the spin independent cross section could be suppressed heavily
\cite{Lu:2016ucn}, therefore $\Delta^0$ can easily escape direct
detection even in the future.

In model (f) and (i), an inert scalar doublet $\eta\sim(1,2,1)$ and
a complex inert scalar triplet $\Delta\sim(1,3,2)$ are
added\cite{Lu:2016dbc}. Naively, we expect that the DM candidate is
$\eta_R^0$ or $\eta_I^0$ in these two models, since $\Delta_R^0$ or
$\Delta_I^0$ cannot play the role of DM candidate solely
if $\eta$ does not exist. However, in these two models,
a mass splitting $\Delta M=|M_{\Delta_R^0}-M_{\Delta_I^0}|$ between
$\Delta_R^0$ and $\Delta_I^0$ exists due to the mixing between
$\eta$ and $\Delta$ \cite{Kajiyama:2013zla}. Specifically speaking,
the $\kappa (\eta^\dag H)^2$ term will induce the mass
splitting between $\eta_R^0$ and $\eta_I^0$. Then the trilinear term
$\mu H^T i \sigma_2 \Delta^\dag \eta$ will induce the mixing between
$\Delta_R^0$ and $\eta_R^0$ for the CP-even scalars, and  mixing
between $\Delta_I^0$ and $\eta_I^0$ for the CP-odd scalars,
resulting a mass splitting between $\Delta_R^0$ and $\Delta_I^0$.
For instance, with $M_{\eta_R^0}=5~\TeV$,
$M_{\eta_I^0}^2=M_{\eta_R^0}^2-2\kappa v^2$, $\kappa=0.5$,
$\mu=1~\TeV$, and $M_{\Delta}=2.8~\TeV$, the mass splitting $\Delta
M=|M_{\Delta_R^0}-M_{\Delta_I^0}|\approx 1~\MeV$.
 Therefore this mass splitting $\Delta M$ is
larger than the DM kinetic energy $\mathcal{O}(100)~\text{keV}$, the
tree level DM-nucleon scattering via $Z$-boson is expected
kinematically forbidden \cite{Law:2013saa}. In this way,
$\Delta_I^0$ (or $\Delta_R^0$ when $\kappa<0$) can escape the direct
detection bound, thus becomes a viable DM candidate. And
$M_{\Delta_R^0/\Delta_I^0}\sim2.8~\TeV$ is preferred to acquire the
correct DM relic density \cite{Araki:2011hm}.

In model (g) and (j), the only scalar DM candidate is
$\Delta\sim(1,3,-2)$, since the other scalar $\eta\sim(1,2,3)$ does
not have neutral component. But the scalar triplet $\Delta$ has
already excluded by the direct detection experiments
\cite{Araki:2011hm}. Therefore, these two models could not provide
viable DM candidate.

Note that the discrete $Z_2^D$ symmetry could be an accidental
symmetry of a broken $U(1)_D$ symmetry \cite{Krauss:1988zc}.
Usually, a SM scalar singlet $\sigma$ is introduced to break
$U(1)_D\to Z_2^D$ spontaneously. Under this extended $U(1)_D$
symmetry, the inert fermion $F$ as well as inert scalars $S_{1,2}$
carry certain $U(1)_D$ charges. While all other ingredients could be
the same as the $Z_2^D$ case, the quartic  term $\kappa (\eta^\dag
H)^2$ for $\eta\sim(1,2,1)$ or $\kappa (\eta^T H)^2$ for
$\eta\sim(1,2,-1)$ is forbidden by the $U(1)_D$ symmetry. The absent
of this quartic term will lead to degenerate masses of $\eta_R^0$
and $\eta_I^0$ in model (b) and (e), therefore they will be excluded
by direct detection in the case of $U(1)_D$ symmetry. Similar for
model (f) and (i), $\eta_R^0$ and $\eta_I^0$ are degenerate, thus
$\Delta_R^0$ and $\Delta_I^0$ are also degenerate. In this way,
model (f) and (i) are also excluded. Meanwhile for model (a), (c) or
(d), (h), mixing between $\phi\sim(1,1,0)/\Delta\sim(1,3,0)$ and
doublet $\eta$ can also lead to a mass splitting between $\eta_R^0$
and $\eta_I^0$. And $\eta_R^0$ is the DM candidate when
$\phi/\Delta$ is heavier than $\eta$.

Last but not least, we give some comments on models with quadruplets
or quintuplets in Table~\ref{tabnew3}. Obviously, model (m), (n),
(q), (r), (s), (v), (w) and (s) have already excluded by direct
detection, since the neutral components in these models have
non-zero hyper-charge and no mass splitting between the real and
imaginary part of the neutral fields could be induced. Model (k) and
(t) are the only two models with viable fermion DM and quadruplets
or quintuplets. For scalar DM, it could be inert triplet or
quintuplet with $Y=0$ as in model (k), (o) and (t). Note that the
quartic term $\kappa(\chi^\dag H)^2$ for $\chi\sim(1,4,1)$ or
$\kappa(\chi^T H)^2$ for $\chi\sim(1,4,-1)$ is allowed by the
$Z_2^D$ symmetry. Analogy to the inert doublet, this quartic will
split the neutral components $\chi_R^0$ and $\chi_I^0$, which makes
$\chi_R^0$ or $\chi_I^0$ a viable DM candidate. A mass splitting
between real and imaginary part of the neutral fields in
triplet/quintuplet, i.e., model (l), (p) and (u), is also possible
due to the mixing between triplet/quintuplet and quadruplet. In this
way, the corresponding $Y\neq0$ triplet/quintuplet can avoid the
tight direct detection bounds as well in the present of
$\chi\sim(1,4,\pm1)$.

\section{Phenomenology}\label{PH}

The natural Dirac seesaw models introduce two additional scalars and
a heavy intermediate fermion, which would lead to rich
phenomenology. In this section we choose model (B) for tree level models
and model (a) for one-loop level models as our benchmark mark models to illustrate the relative phenomenon. We briefly highlight some important aspects, although a detailed research on phenomenology of other specific models is quite necessary.

\subsection{Flavor Constraints}

\begin{figure}[!htbp]
\begin{center}
\includegraphics[width=0.45\linewidth]{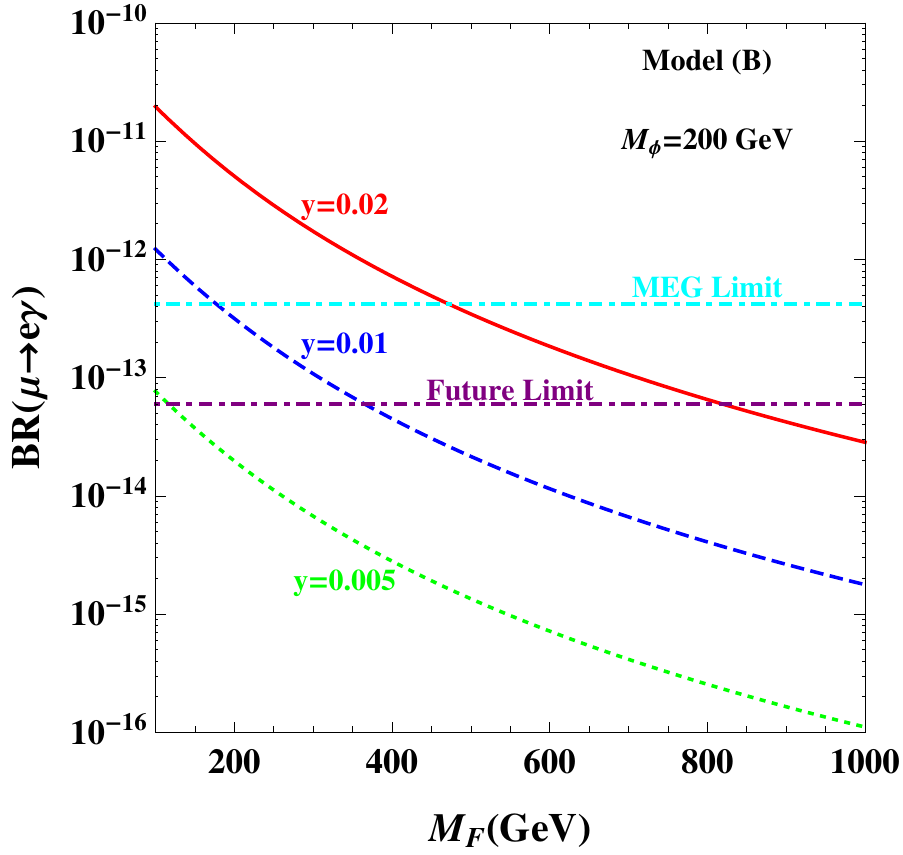}
\includegraphics[width=0.45\linewidth]{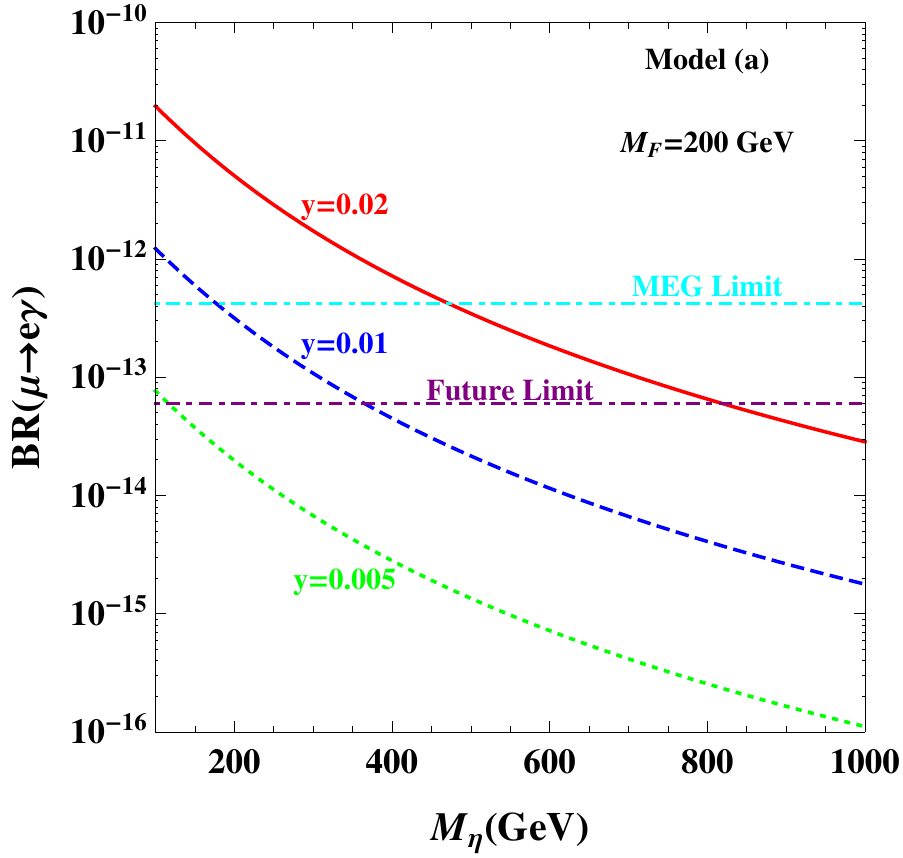}
\end{center}
\caption{BR($\mu\to e\gamma$) as a function of mass of heavy intermediate particle for tree level model (B) and one-loop level model (a). Here, we assume an universal Yukawa coupling $|y_1^{ij}|=y$ and degenerate masses for the three generation of heavy fermion $F$ for simplicity. We have $S_1=\phi$ in model (B) and $S_1=\eta$ in model (a), respectively.
\label{Fig:LFV}}
\end{figure}

The existence of Yukawa coupling $y_1 \overline{F_R} L_L S_1$ will
induce lepton flavor violation (LFV) processes.  Here, we take the current most stringent bound
BR$(\mu\to e \gamma)<4.2\times10^{-13}$ \cite{Adam:2013mnn} and future limit BR$(\mu\to e \gamma)<6\times10^{-14}$ \cite{Baldini:2013ke} to
illustrate, and more discussion on other LFV processes can be found
in Refs. \cite{Ma:2001mr,Liao:2009fm,Lindner:2016bgg}. The general analytical
expression for BR$(\mu\to e \gamma)$ is given by
\begin{equation}\label{BR}
\text{BR}(\mu\to e\gamma)=\frac{3\alpha}{64\pi G_F^2}\left|\sum_{i=1}^3
\frac{y_1^{ie*}y_1^{i\mu}}{M_{S_1}^2}\left[Q_{F_i}F_{1}\!\left(\frac{M_{F_i}^2}{M_{S_1}^2}
\right)+Q_{S_1}F_{2}\!\left(\frac{M_{F_i}^2}{M_{S_1}^2}\right)\right]\right|^2,
\end{equation}
where the loop functions $F_{i}(x)$ are \cite{Ding:2014nga}
\begin{eqnarray}
F_1(x)&=&\frac{2+3x-6x^2+x^3+6x\ln x}{6(1-x)^4},\\
F_2(x)&=&\frac{1-6x+3x^2+2x^3-6x^2\ln x}{6(1-x)^4}.
\end{eqnarray}
Here, $Q_{F_i}$ and $Q_{S_1}$ denote the electric charge of charged components in $F_i$ and $S_1$, respectively. More specifically, we have $Q_{F_i}=1$, $Q_{S_1}=0$ in model (B) and $Q_{F_i}=0$, $Q_{S_1}=1$ in model (a). In Fig.~\ref{Fig:LFV}, we depict the predicted value of $\text{BR}(\mu\to e\gamma)$ as a function of charged particle mass in this two models with an universal Yukawa coupling $|y_1^{ij}|=y$ and degenerate masses for the three generation of heavy fermion $F$. Constraints on the Yukawa coupling for this two models are similar. And for both tree-level model (B) and purely radiative model (a), the tight constraints from LFV usually requires that the corresponding Yukawa coupling $|y_1|\lesssim0.01$ with $F$ and $S_1$ around electroweak
scale~\cite{Liao:2009fm}.

For tree level models, the tight upper bound on branching ratios of LFV could be
transformed into a lower bound on $\langle S_1 \rangle$\cite{Ding:2014nga}.
From the expression of neutrino mass in Eq.~\ref{mv-tree}, it is estimated
that $y_1\simeq m_\nu^{1/2} M_F^{1/2}/\langle S_1 \rangle$ by
assuming $y_1\simeq y_2$ and $\langle S_1\rangle\simeq\langle S_2
\rangle$. Plugging this estimation into Eq.~\ref{BR}, one easily
derives
\begin{eqnarray}
M_{S_1} \langle S_1 \rangle &\gtrsim& \left(\frac{3\alpha m_\nu^2
M_F^2 }{64\pi G_F^2 \text{BR}(\mu\to e\gamma)}\left|\sum_{i=1}^3
F_1\left(\frac{M_{F_i}^2}{M_{S_1}^2}\right)\right|^2\right)^{1/4},\\\nonumber
&\approx&\left[\frac{1\times10^{-13}}{\text{BR}(\mu\to
e\gamma)}\left(\frac{M_{F}}{100~\GeV}\right)^2\right]^{1/4}\hspace{-1em}\times
600~\GeV\cdot\MeV,
\end{eqnarray}
where we have assume $m_\nu\sim0.1~\eV$ and $\sum F_1(x)\sim0.1$
in the numerical estimation. For electroweak scale intermediate
fermion $M_{F}\sim200~\GeV$, current limits on BR($\mu \to e\gamma$)
requires that $M_{S_1} \langle S_1 \rangle \gtrsim
600~\GeV\cdot\MeV$. Thus, the VEVs of scalars $S_{1,2}$ are expected
to be larger than $\mathcal{O}(\MeV)$ when the mass of $S_{1,2}$ is
around electroweak scale as well.

The contribution to the anomalous magnetic moment of $\mu$ can be obtained as a by-product of the above calculation of LFV
\begin{equation}
\Delta a_\mu = \sum_i^3 \frac{|y_1^{i\mu}|^2}{16\pi^2}\frac{M_\mu^2}{M_{S_1}^2}
\left[Q_{F_i}F_{1}\!\left(\frac{M_{F_i}^2}{M_{S_1}^2}
\right)+Q_{S_1}F_{2}\!\left(\frac{M_{F_i}^2}{M_{S_1}^2}\right)\right].
\end{equation}
Under constraints from LFV, the predicted value of $\Delta a_\mu$ is $4\times10^{-14}$ for an universal Yukawa coupling $y_1\sim0.01$ and both $F$ and $S_1$ around electroweak scale, which is clearly too small to interpret the observed discrepancy $\Delta a_\mu=(2.39\pm0.79)\times10^{-9}$ \cite{Bennett:2006fi}.

Another tight constraint comes from electric dipole moments (EDM) of electron, which requires $|d_e|<8.7\times10^{-29}~e$-cm \cite{Baron:2013eja}. In all the current Dirac neutrino models, the only new interactions for lepton doublet $L_L$ is the Yukawa coupling $y_1 \overline{F_R} L_L S_1$, which can not give large contributions to EDM at one-loop level \cite{Barger:1996jc,Fukuyama:2012np}. Actually, the contribution of above Yukawa coupling $y_1 \overline{F_R} L_L S_1$ to electron EDM first appears at two-loop level (see Fig.~4 of Ref.~\cite{Borah:2016zbd}). Considering constraints from LFV, a naive estimation for the order of magnitude gives \cite{Borah:2016zbd}
\begin{equation}
d_e \sim \frac{M_e\text{Im}(y_1^2 \lambda)}{(16\pi^2)^2 M_{S_1}^2}\sim10^{-31}~e\text{-cm},
\end{equation}
with $M_{S_1}\sim200~\GeV$, $y_1\sim0.01$, and Im$(\lambda)\sim0.1$. Here, $\lambda$ is the coefficient of the quartic coupling $S_1^\dag S_1 H^\dag H$. Therefore, the contribution of the new Yukawa coupling $y_1 \overline{F_R} L_L S_1$ to electron EDM is about two to three orders of magnitude lower than current limit with the above parameters.

\subsection{Leptogenesis}

Within Majorana seesaw models, the observed baryon asymmetry can be
explained via conventional leptogenesis \cite{Fukugita:1986hr},
where the lepton number violation plays an essential role.
Obviously, no lepton asymmetry is generated in Dirac seesaw models
because the lepton number is conserved. However, the leptogenesis
can still be accomplished in Dirac neutrino
models \cite{Dick:1999je}, due to the fact that the sphaleron
processes do not have direct effect on right-handed fields.
Therefore, if an equal but opposite amount of lepton asymmetry in
the left- and right-handed sectors is created, the lepton asymmetry
in the left-handed sector can be converted into a net baryon
asymmetry via sphaleron processes, as long as the effective Dirac
Yukawa couplings are small enough to prevent the lepton asymmetry
from equilibration before the electroweak phase transition. Detailed
studies on Dirac leptogenesis can be found in Ref.
\cite{Murayama:2002je}. For the the models we discussed, the
required lepton asymmetry in left- and right-handed sectors arises
from the decays of the heavy intermediate fermion $F$ into $L_L S_1$
and $\nu_R S_2$.

\begin{figure}
\begin{center}
\includegraphics[width=0.45\linewidth]{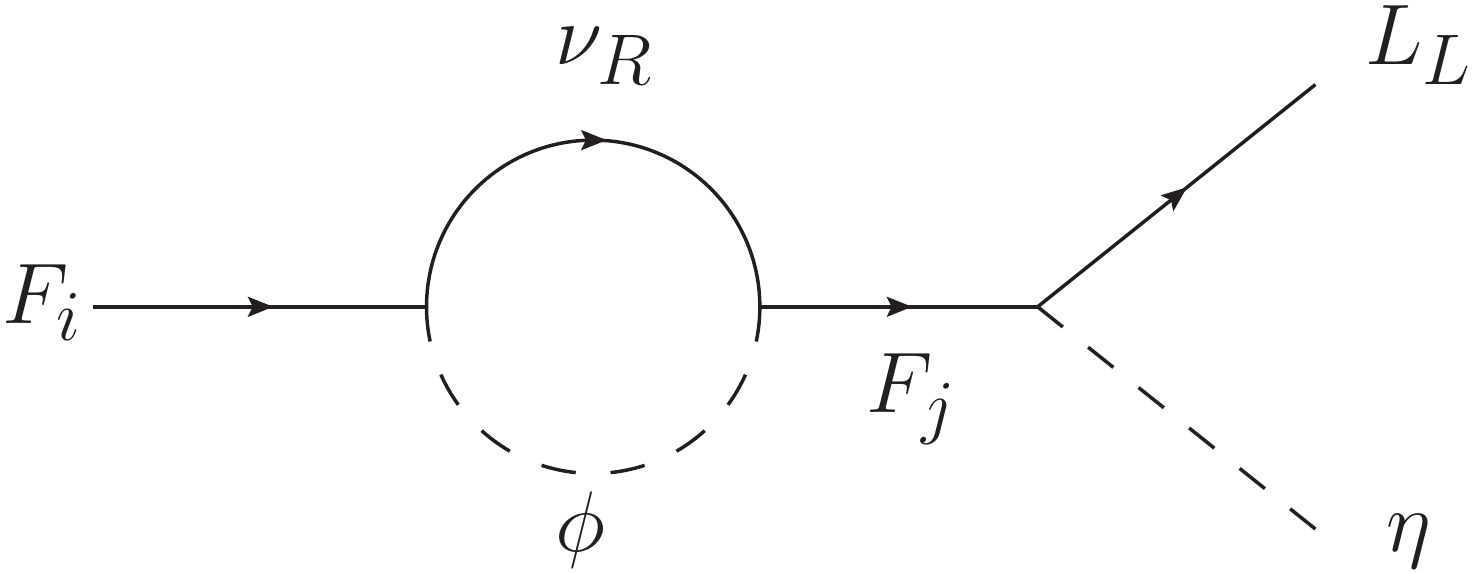}
\end{center}
\caption{Heavy Dirac fermions $F_i$ decay into left-handed leptons at one-loop level.
\label{Fig:LG}}
\end{figure}

For the tree-level model (B), it is possible to generate the baryon asymmetry via resonant leptogenesis with nearly degenerate $F_i$ around $\TeV$-scale \cite{Pilaftsis:2003gt}. For simplicity, we consider the canonical thermal leptogenesis in the one-loop model (a), where very heavy $F_i$ is needed.
The heavy Dirac fermion $F_i$ has two decay modes:
$F_i\to L_L \eta$ and $F_i\to \nu_R \phi$, and the corresponding decay widths at tree level are
\begin{eqnarray}\label{GF1}
\Gamma(F_i\to L_L \eta)=\Gamma(F^C_i\to L_L^C \eta^*)&=& \frac{M_{F_i}}{16\pi}(y_1^\dag y_1)_{ii},\\ \label{GF2}
\Gamma(F_i\to \nu_R \phi)=\Gamma(F^C_i\to \nu_R^C \phi)~~&=& \frac{M_{F_i}}{32\pi}(y_2^\dag y_2)_{ii},
\end{eqnarray}
in the limit of $M_{\eta,\phi}\ll M_{F_i}$. As shown in Fig.~\ref{Fig:LG}, the required lepton
asymmetry in the left-handed sector arise at one-loop level and is calculated as \cite{Gu:2007ug}
\begin{eqnarray}\nonumber
\epsilon_{F_i} &=& \frac{\Gamma(F_i\to L_L \eta)-\Gamma(F^C_i\to L_L^C \eta^*)}{\Gamma_{F_i}} \\
&=&\frac{1}{8\pi} \frac{1}{(y_1^\dag y_1)_{ii}+\frac{1}{2}(y_2^\dag y_2)_{ii}}\sum_{j\neq i} \text{Im}\left[(y^\dag_1y_1)_{ij}(y^\dag_2y_2)_{ji}\right]\frac{M_{F_i}M_{F_j}}{M_{F_i}^2-M_{F_j}^2},
\end{eqnarray}
where the total decay width is $\Gamma_{F_i}=[(y_1^\dag y_1)_{ii}+(y_2^\dag y_2)_{ii}/2]M_{F_i}/(16\pi)$. Provided that $M_{F_1}\ll M_{F_{2,3}}$, then the final left-handed sector lepton asymmetry is dominantly determined by the decays of $F_1$:
\begin{equation}
\epsilon_{F_1}\approx
-\frac{1}{8\pi}\frac{1}{(y^\dag_1y_1)_{11}+\frac{1}{2}(y^\dag_2y_2)_{11}}
\sum_{j\neq1}\frac{M_{F_1}}{M_{F_j}}\text{Im}\left[(y^\dag_1
y_1)_{1j}(y^\dag_2y_2)_{j1}\right],
\end{equation}
We further take $y_1=y_2$ for illustration, then the lepton asymmetry $\epsilon_{F_1}$ can be simplified as
\begin{equation}
\epsilon_{F_1}\simeq -\frac{1}{24\pi}\frac{1}{(y^\dag_1y_1)_{11}}
\sum_{j\neq1}\frac{M_{F_1}}{M_{F_j}}\text{Im}\left[(y^\dag_1
y_1)_{1j}^2\right].
\end{equation}
With the assumption $y_1=y_2$, an upper bound on $\epsilon_L$ can be
deduced after considering the radiative neutrino masses in
Eq.~\ref{mv:loop1} \cite{Gu:2007ug}
\begin{equation}
|\epsilon_{F_1}|\lesssim \frac{4\pi M_{F_1}
m_{3} |\sin \delta|}{3 \sin2\theta \left|M_{\eta}^2 \ln \frac{M_\eta^2}{M_{F_1}^2}
-M_{\phi}^2 \ln \frac{M_\phi^2}{M_{F_1}^2}\right|},
\end{equation}
with $m_3$ the heaviest neutrino mass and $\delta$ the Dirac phase.
Setting $M_{F_1}=10^7~\GeV$, $M_\phi=60~\GeV$, $m_3=0.1~\eV$, $M_{\eta}=200~\GeV$,
$\theta=0.01$ and $\sin \delta=-1$, we obtain
$\epsilon_{F_1}\simeq-2.7\times10^{-7}$. Then after the sphaleron processes,
the desired baryon asymmetry
\begin{equation}
Y_B=\frac{n_B-n_{\bar{B}}}{s}=-\frac{28}{79}\frac{n_{L}}{s}
\simeq-\frac{28}{79}\epsilon_{F_1}\frac{n_{F_1}^{\text{eq}}}{s}\Big|_{T=M_{F_1}}
\simeq -\frac{\epsilon_{F_1}}{15g_{*}}\approx1.7\times10^{-10}
\end{equation}
with $g_{*}=106.75$ is obtained to explain the observed
baryon asymmetry \cite{Ade:2015xua}.
\begin{figure}
\begin{center}
\includegraphics[width=0.45\linewidth]{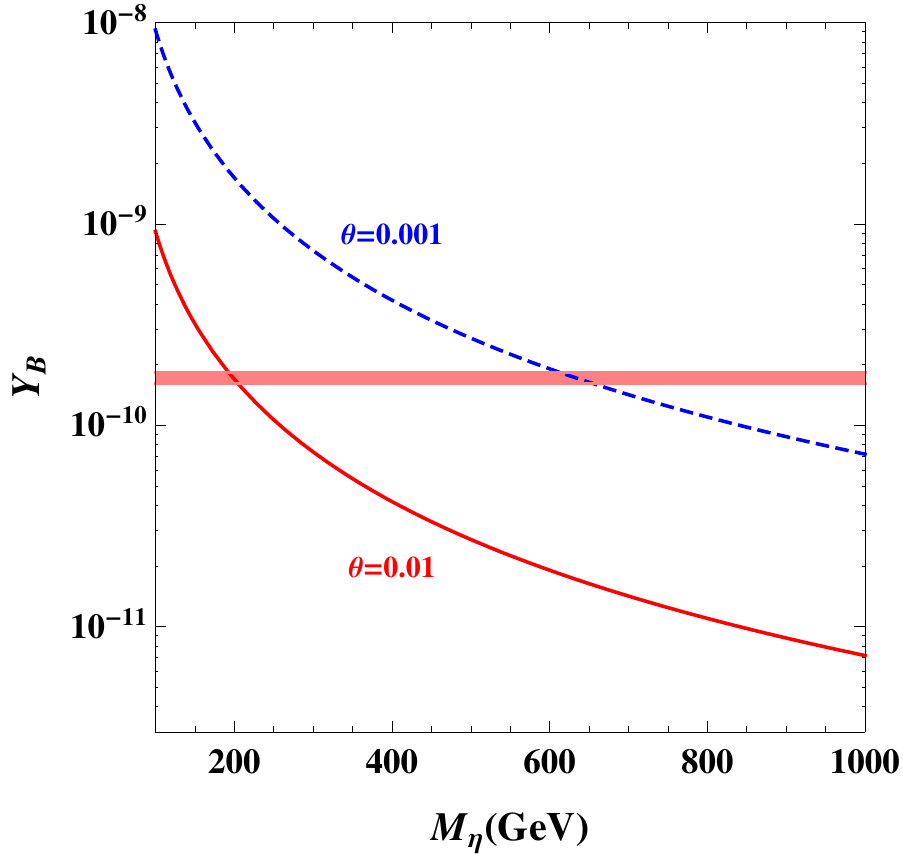}
\end{center}
\caption{$Y_B$ as a function of $M_\eta$ for $\theta=0.01,0.001$. The pink band corresponds to $1\sigma$ range of the observed value in Ref.~\cite{Ade:2015xua}.
\label{Fig:YB}}
\end{figure}
In Fig.~\ref{Fig:YB}, we show the value of $Y_B$ as a
function of $M_\eta$ for $\theta=0.01,0.001$. With other parameters fixed, the larger $M_\eta$ is,
the smaller the $\theta$ is required to obtain the observed value of $Y_B$. Meanwhile, the decays of
$F_1$ should be out of equilibrium, which requires that
\begin{equation}
\Gamma_{F_1}\lesssim H(T)\Big|_{T=M_{F_1}},~\text{with}~H(T)=\left(\frac{8\pi^2 g_{*}}{90}\right)^{\frac{1}{2}}\frac{T^2}{M_{\text{Pl}}},
\end{equation}
where $M_{\text{Pl}}=1.22\times10^{19}~\GeV$. With the assumption $y_1=y_2$, $M_{F_1}\sim10^7~\GeV$ and Eq.~\ref{GF1},\ref{GF2},
the above condition indicates that the Yukawa coupling $y_1$ should satisfy
\begin{equation}
(y_1^\dag y_1)_{11}\lesssim \left(\frac{2^{10}\pi^5g_{*}}{5*3^4}\right)^{\frac{1}{2}}\frac{M_{F_1}}{M_{\text{Pl}}}\sim10^{-10}.
\end{equation}

\subsection{Dark Matter}

In the two benchmark model we studied,
there is no DM candidate in the tree level model (B). Meanwhile,
for the one-loop model (a), there are viable DM candidate $\phi$ or $\eta_{R,I}^0$.
In this work, we consider the case of $M_\phi<M_\eta$ with small mixing angle $\theta\lesssim0.01$, thus the DM candidate is dominantly determined by $\phi$. The relic density of $\phi$ is mostly determined by the quartic coupling $\lambda_\phi \phi^2 H^\dag H$, and the analytic expression is given by \cite{Gondolo:1990dk}
\begin{equation}
\Omega_\phi h^2 = \frac{1.07\times10^9 \GeV^{-1}}{\sqrt{g_*} M_{\text{Pl}}J(x_f)},
\end{equation}
where the function $J(x_f)$ is
\begin{equation}
J(x_f)=\int_{x_f}^\infty\frac{\langle \sigma v_{\text{rel}}\rangle (x)}{x^2}dx.
\end{equation}
And the freeze-out parameter $x_f=M_\phi/T_f$ is acquired by numerically solving
\begin{equation}
x_f = \ln \left(\frac{0.038 M_{\text{Pl}} M_\phi \langle \sigma v_{\text{rel}}\rangle (x_f)}{\sqrt{g_* x_f}}\right).
\end{equation}
\begin{figure}
\begin{center}
\includegraphics[width=0.45\linewidth]{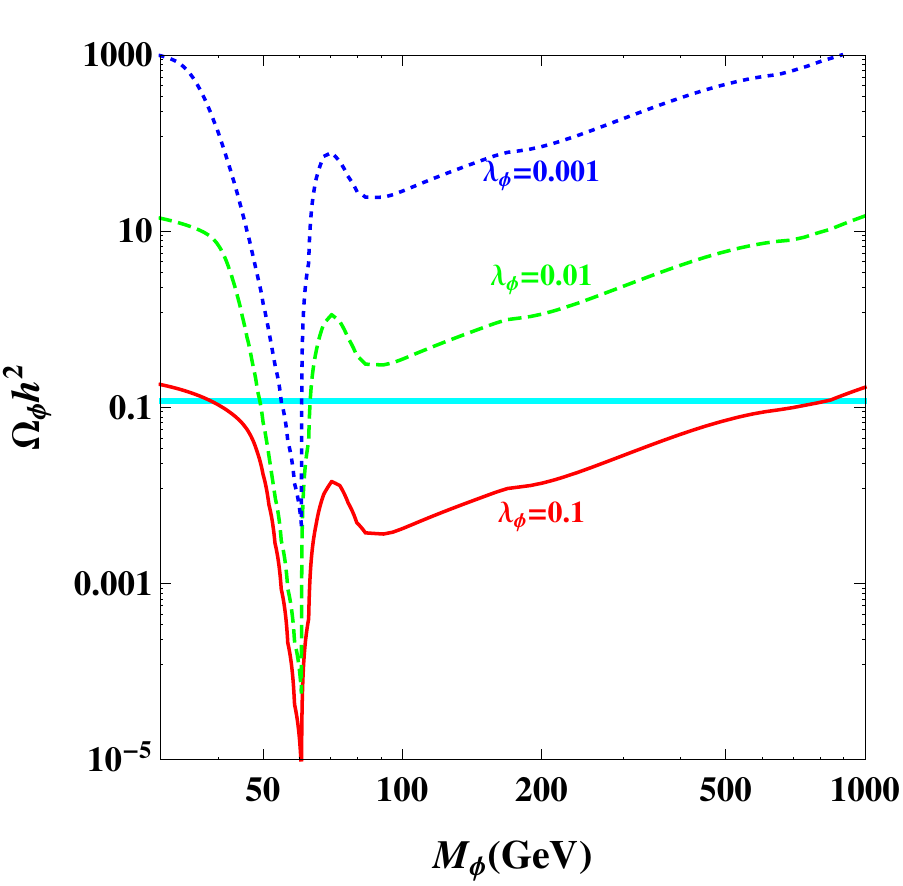}
\end{center}
\caption{Relic density $\Omega_\phi h^2$ as a function of $M_{\phi}$ for $\lambda_{\phi}=0.001$, $0.01$, and $0.1$. The cyan band corresponds to the observed DM relic density \cite{Ade:2015xua}.
\label{Fig:RC}}
\end{figure}

As pointed out by Ref.~\cite{Cline:2012hg}, the QCD corrections for quarks in the final state, as well as three- and four-body final states from virtual gauge boson decays are important for total DM annihilation cross section. Following Ref.~\cite{Cline:2013gha}, we rewrite the annihilation cross section into all SM particles  except $h$ as
\begin{equation}
\sigma v_\text{rel} = \frac{8 \lambda_\phi^2 v^2}{\sqrt{s}} \frac{\Gamma_h(\sqrt{s})}{(s-M_h^2)^2+M_h^2\Gamma_h^2(M_h)},
\end{equation}
where $v=246~\GeV$ and the tabulated accurate Higgs boson width as a function of invariant mass $\Gamma_h(\sqrt{s})$ can be found in Ref.~\cite{Heinemeyer:2013tqa}. For light DM $M_\phi<M_h/2$, the decay width $\Gamma_h(M_h)$ in the denominator should add the contribution of Higgs invisible decay $h\to \phi\phi$. Meanwhile, for heavy DM $M_\phi > M_h$, the extra contribution from $\phi\phi\to hh$ has also to be supplemented. But above $M_\phi>150~\GeV$, we should use the tree-level expressions in Appendix B, since the loop corrections are overestimated \cite{Cline:2013gha}. The thermal average cross section is then carried out via
\begin{equation}
\langle \sigma v_\text{rel} \rangle (x) = \frac{x}{16 M_\phi^5 K_2^2(x)}\int_{4M_\phi^2}^\infty
\sqrt{s-4M_\phi^2} s K_1\left(\frac{x\sqrt{s}}{M_\phi}\right) \sigma v_\text{rel} ds,
\end{equation}
where $K_{1,2}(x)$ are modified Bessel functions of the second kind. In Fig.~\ref{Fig:RC}, we show the relic density $\Omega_\phi h^2$ as a function of $M_{\phi}$ for $\lambda_{\phi}=0.001$, $0.01$, and $0.1$. The correct relic density can be obtained in the low-mass region $M_\phi<M_h/2$ and high-mass region $M_\phi>M_h/2$ for fixed value of $\lambda_\phi$.

\begin{figure}
\begin{center}
\includegraphics[width=0.45\linewidth]{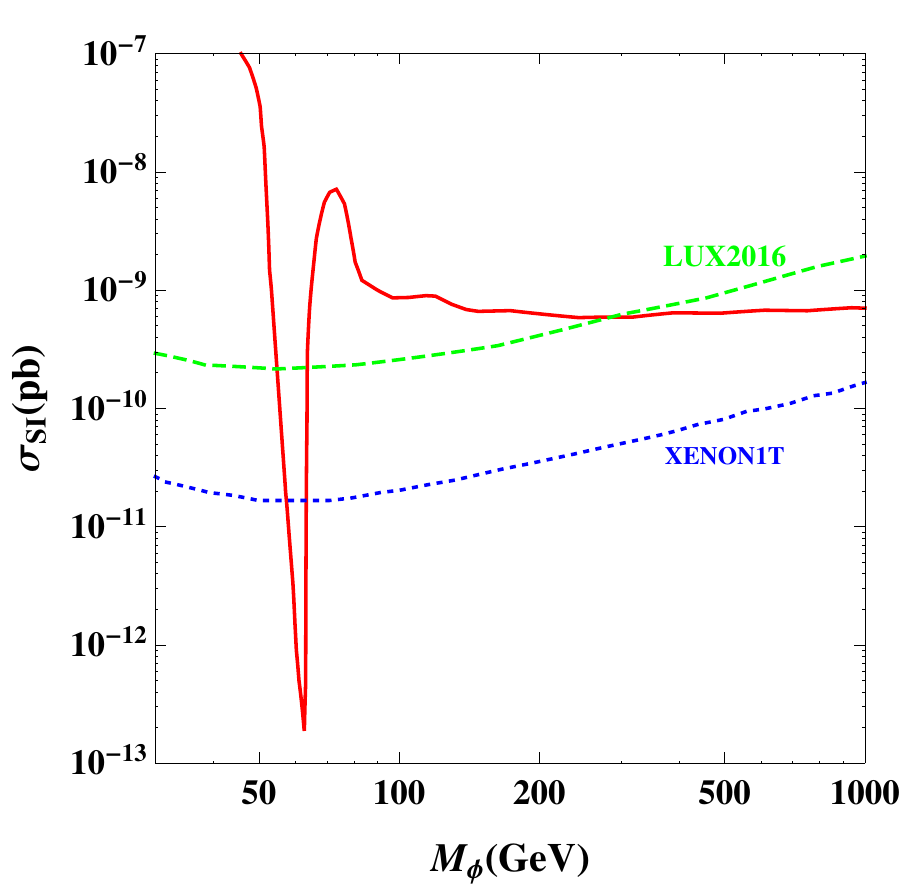}
\end{center}
\caption{The spin-independent DM-nucleon cross section $\sigma_\text{SI}$ as a function of $M_{\phi}$. The green and blue lines correspond to LUX2016~\cite{Akerib:2013tjd} and XENON1T~\cite{Aprile:2012zx} limits.
\label{Fig:DD}}
\end{figure}

Then we consider possible constraints from DM direct detection. The cross section for spin independent DM-nucleon is
\begin{equation}
\sigma_{\text{SI}} = \frac{\lambda_\phi^2 f_N^2 \mu^2 m_N^2}{\pi M_h^4 M_\phi^2},
\end{equation}
where $m_N=(m_p+m_n)/2=939~\MeV$ is the averaged nucleon mass, $f_N=0.3$ is the matrix element, and $\mu=m_N M_\phi/(m_N+M_\phi)$ is the DM-nucleon reduced mass. Provided $\phi$ accounting for 100\% of DM, the predicted value of $\sigma_\text{SI}$ is presented in Fig.~\ref{Fig:DD}. In the current simple scenario we considered, it is clear that the only possible region to escape tight direct detection constraints is around the Higgs mass resonance, i.e., $M_\phi\approx M_h/2$. Thus, the choice of $M_\phi=60~\GeV$ in this work is safe to avoid direct detection constraints.

\begin{figure}
\begin{center}
\includegraphics[width=0.45\linewidth]{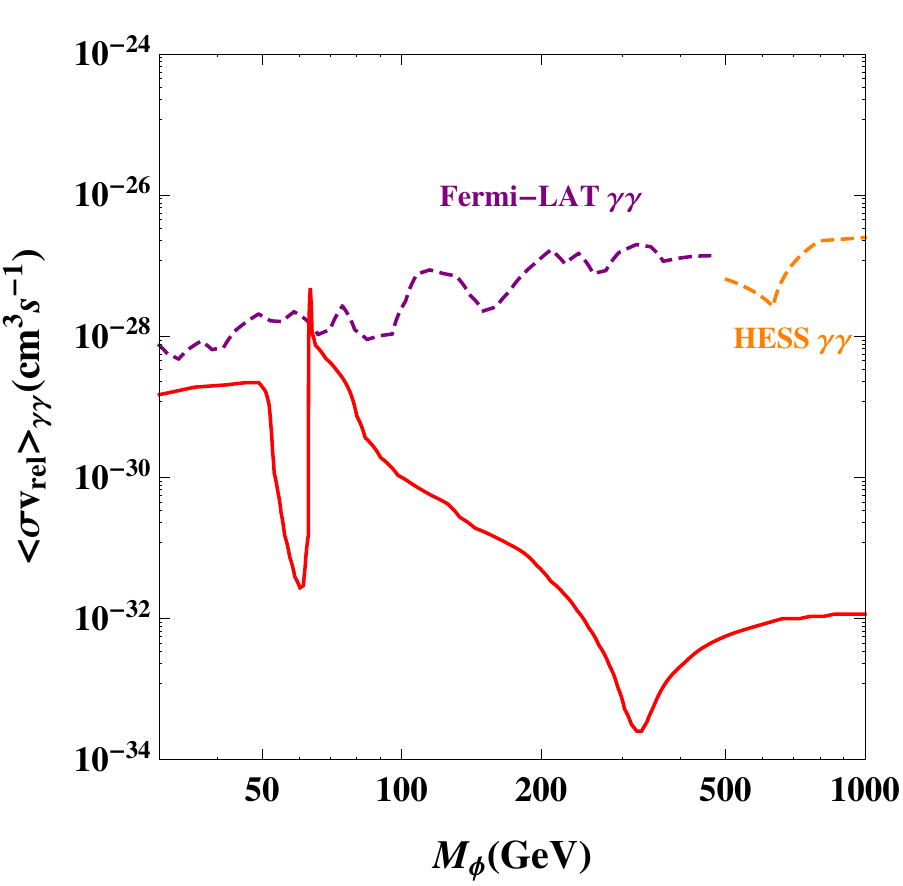}
\includegraphics[width=0.45\linewidth]{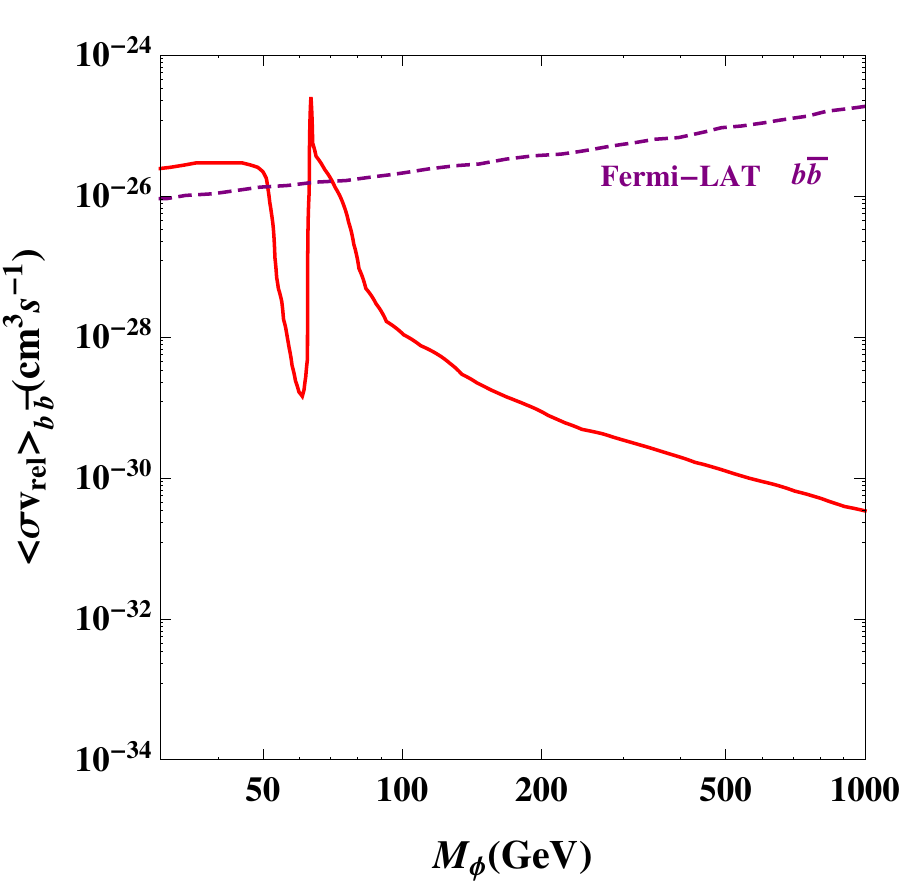}
\end{center}
\caption{The velocity-averaged annihilation cross section times relative velocity $\langle\sigma v_{\text{rel}} \rangle$ into $\gamma\gamma$ (left) and $b\bar{b}$ (right).
\label{Fig:ID}}
\end{figure}

There are also possible constraints from indirect detection. In Fig.~\ref{Fig:ID}, we depict the predictions for $\langle\sigma v_{\text{rel}} \rangle_{\gamma\gamma,b\bar{b}}$, as well as the observed limits from Fermi-LAT~\cite{Ackermann:2015zua} and H.E.S.S.~\cite{Abramowski:2013ax}. In the $\gamma\gamma$ final state, only a tiny mass region $M_h/2\lesssim M_{\phi}$ is excluded. Meanwhile, in the $b\bar{b}$ final state, two mass region $M_\phi<51~\GeV$ and $M_h/2\lesssim M_{\phi}<70~\GeV$ are excluded.

\begin{figure}
\begin{center}
\includegraphics[width=0.45\linewidth]{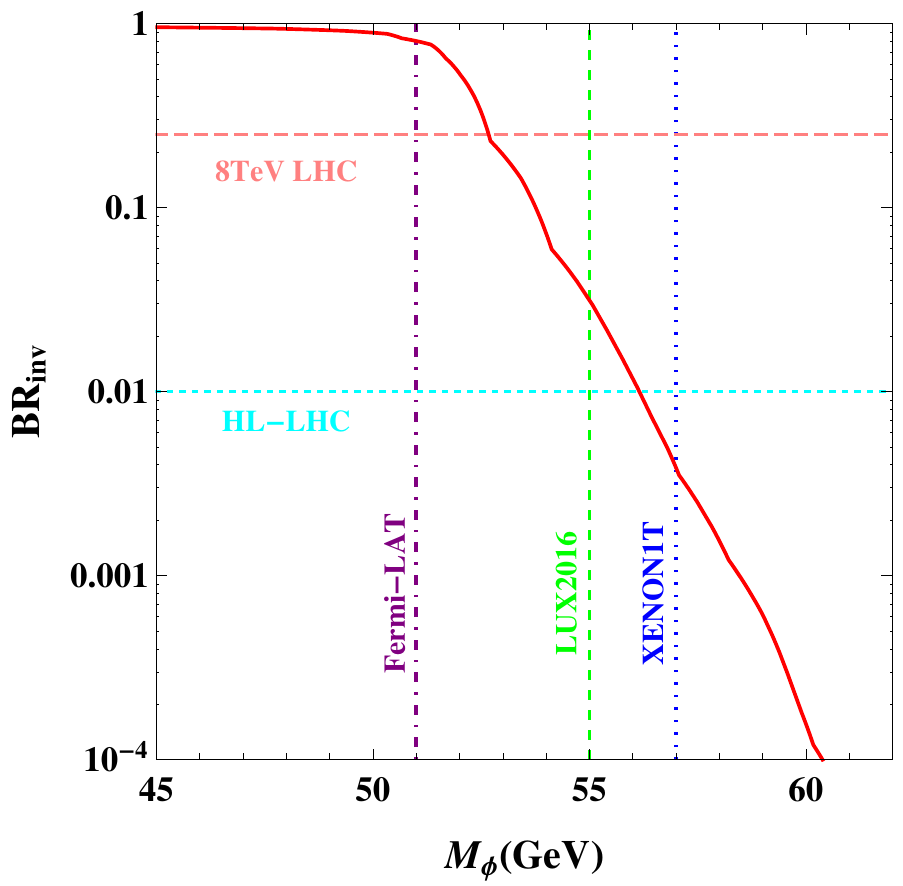}
\end{center}
\caption{Branching ratio of Higgs invisible decay BR$_\text{inv}$ as a function of $M_{\phi}$.
\label{Fig:Inv}}
\end{figure}

Last but not least, the SM Higgs $h$ will decay into DM pair in the low mass region $M_\phi<M_h/2$, which will induce Higgs invisible decay at colliders. The corresponding decay width is
\begin{equation}
\Gamma(h\to \phi\phi)= \frac{\lambda_\phi^2 v^2}{8\pi M_h^2}\sqrt{M_h^2-4M_\phi^2},
\end{equation}
and the invisible branching ratio is $\text{BR}_\text{inv}=\Gamma(h\to \phi\phi)/(\Gamma(h\to \phi\phi)+\Gamma_\text{SM})$, where $\Gamma_\text{SM}=4.07~\MeV$ for $M_h=125~\GeV$ \cite{Heinemeyer:2013tqa}. In Fig.~\ref{Fig:Inv}, we show BR$_\text{inv}$ as a function of $M_{\phi}$ in the low mass region. The 8 TeV LHC limit, i.e., BR$_\text{inv}\lesssim0.25$,  comes from the fitting results of Higgs visible decay \cite{Khachatryan:2016vau}. And according to Ref.~\cite{Bernaciak:2014pna}, the HL-LHC might probe BR$_\text{inv}\sim0.02$ in the weak boson fusion channel. The 8 TeV LHC has excluded $M_\phi\lesssim53~\GeV$, which is less stringent than the LUX2016 limit. Meanwhile, the HL-LHC will be capable of excluding $M_\phi<56~\GeV$, which will be less stringent than the XENON1T.

\begin{figure}
\begin{center}
\includegraphics[width=0.45\linewidth]{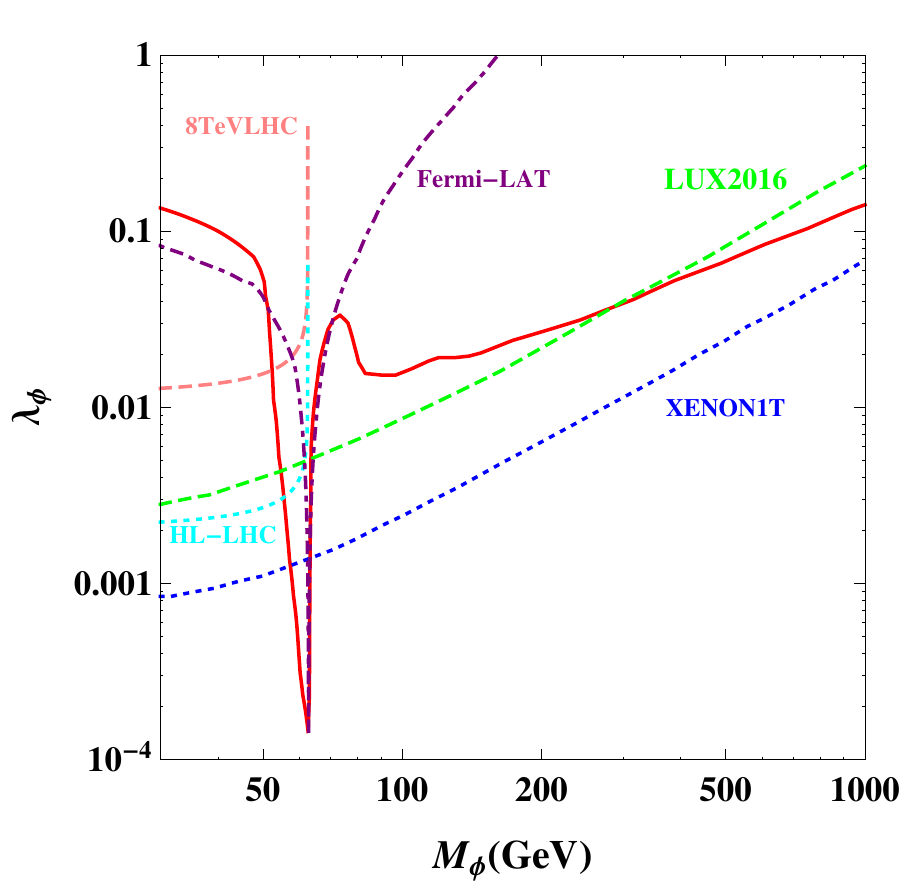}
\end{center}
\caption{Allowed parameter space in the $\lambda_\phi-M_\phi$ plane.
\label{Fig:lam}}
\end{figure}

In summary, we show the allowed parameter space in the $\lambda_\phi-M_\phi$ plane in Fig.~\ref{Fig:lam}, with the constraints from relic density, direct detection, indirect detection and Higgs invisible decay.
Apparently, the only allowed mass region is a narrow one being close to $M_\phi\lesssim M_{h}/2$.

\subsection{LHC Signature}

\begin{figure}
\begin{center}
\includegraphics[width=0.45\linewidth]{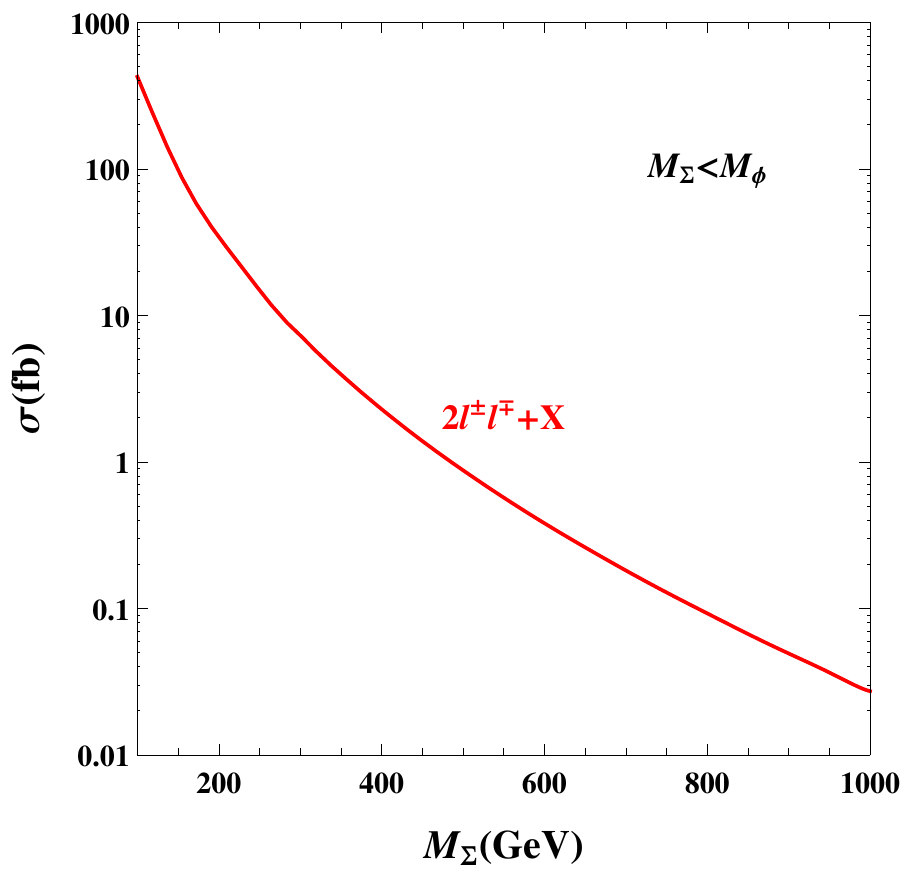}
\includegraphics[width=0.45\linewidth]{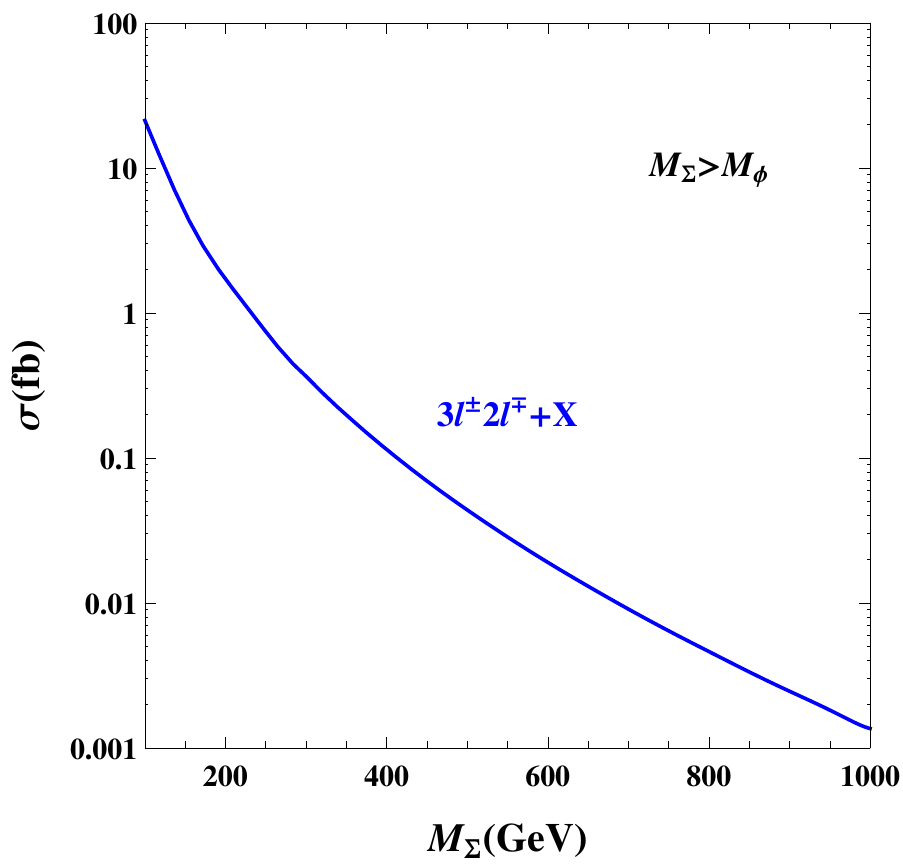}
\end{center}
\caption{Cross section of trilepton signature $2\ell^\pm\ell^\mp+X$ (left) and five-lepton signature $3\ell^\pm+2\ell^\mp+X$ (right) at $14~\TeV$ LHC.
\label{Fig:MBS}}
\end{figure}

Finally we briefly discuss possible LHC signatures. The newly
introduced particles in $F,S_{1,2}$ can be pair/associated produced
via Drell-Yan processes as long as they have non-zero gauge
couplings. Then decays of new particles in $F,S_{1,2}$ will usually
lead to multi-lepton signatures at LHC~\cite{delAguila:2008hw}.
Since production cross section as well as the decay properties of
new particles are model dependent, we take model (B) and model (a)
for illustration here. Detailed study and simulation on specific
models at LHC are highly encouraged to perform. First, for tree
level model (B), a fermion doublet
$F\equiv\Sigma=(\Sigma^0,\Sigma^-)^T\sim(1,2,1)$ and a scalar
singlet $\phi\sim(1,1,0)$ are introduced. Hence, in model (B), only
the fermion doublet $\Sigma$ can be largely produced at LHC via
Drell-Yan processes
\begin{equation}
pp\to \Sigma^+\Sigma^-, \Sigma^0\Sigma^0, \Sigma^\pm\Sigma^0.
\end{equation}
The decay channels of the fermion doublet $F$ are, $\Sigma^0 \to
\ell^-W^+,\nu Z, \nu h$ and $\Sigma^-\to \ell^- Z,\ell^- h, \nu
W^-$. And if $M_\phi<M_\Sigma$, new decay channels as $\Sigma^0\to
\nu \phi$ and $\Sigma^-\to \ell^-\phi$ with $\phi\to W^+W^-, ZZ,hh$
are also possible. Note that in Dirac neutrino mass models, there is
no lepton number violation decays of $\Sigma^0$. Thus with $W^\pm,Z$
decaying leptoniclly, multilepton signatures can be generated. For
$M_{\Sigma}<M_{\phi}$, ATLAS has performed an analysis on the
signatures with three or more leptons based on $\Sigma^\pm\to
\ell^\pm Z\to \ell^\pm \ell^+\ell^-$, and $M_{\Sigma}$ in the range
$114-176~\GeV$ has been excluded \cite{Aad:2015dha}. The cross
section of the inclusive trilepton signature $2\ell^\pm\ell^\mp+X$
is shown in left panel of Fig.~\ref{Fig:MBS}. For
$M_{\Sigma}>M_{\phi}$, the new decay channel $\Sigma^\pm \to
\ell^\pm \phi \to \ell^\pm ZZ\to3\ell^\pm2\ell^\mp$ will lead to
signatures with five or more leptons. And the cross section of this
inclusive five-lepton signature is shown in right panel of Fig.~\ref{Fig:MBS}.

\begin{figure}
\begin{center}
\includegraphics[width=0.45\linewidth]{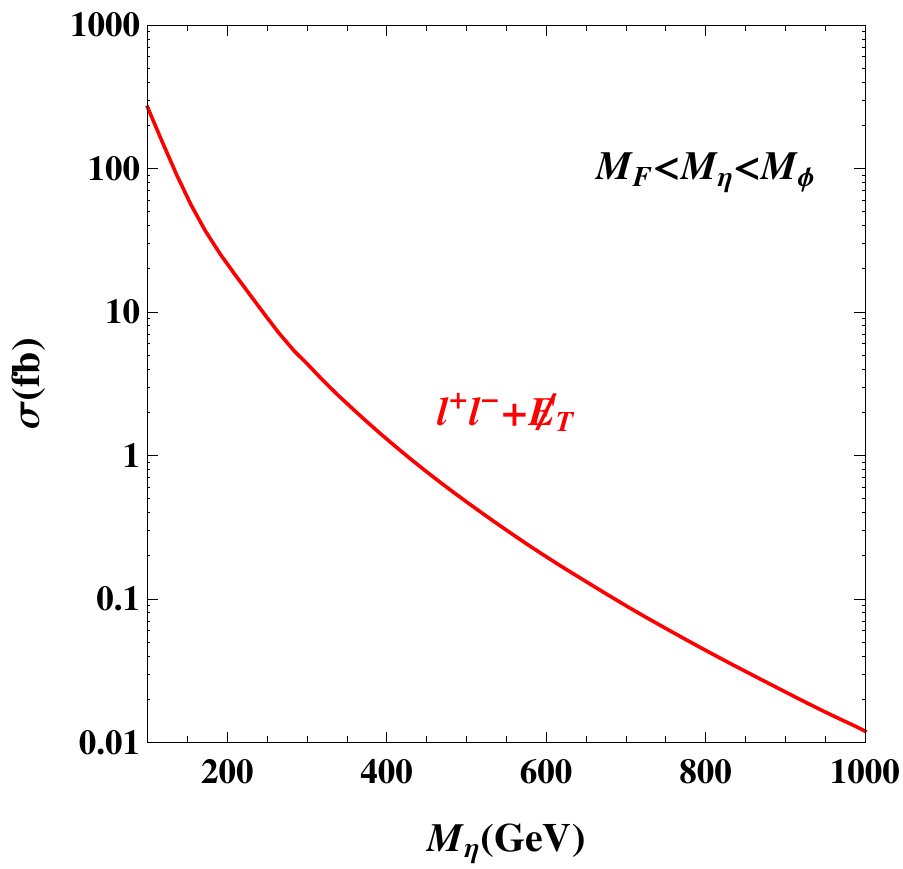}
\includegraphics[width=0.45\linewidth]{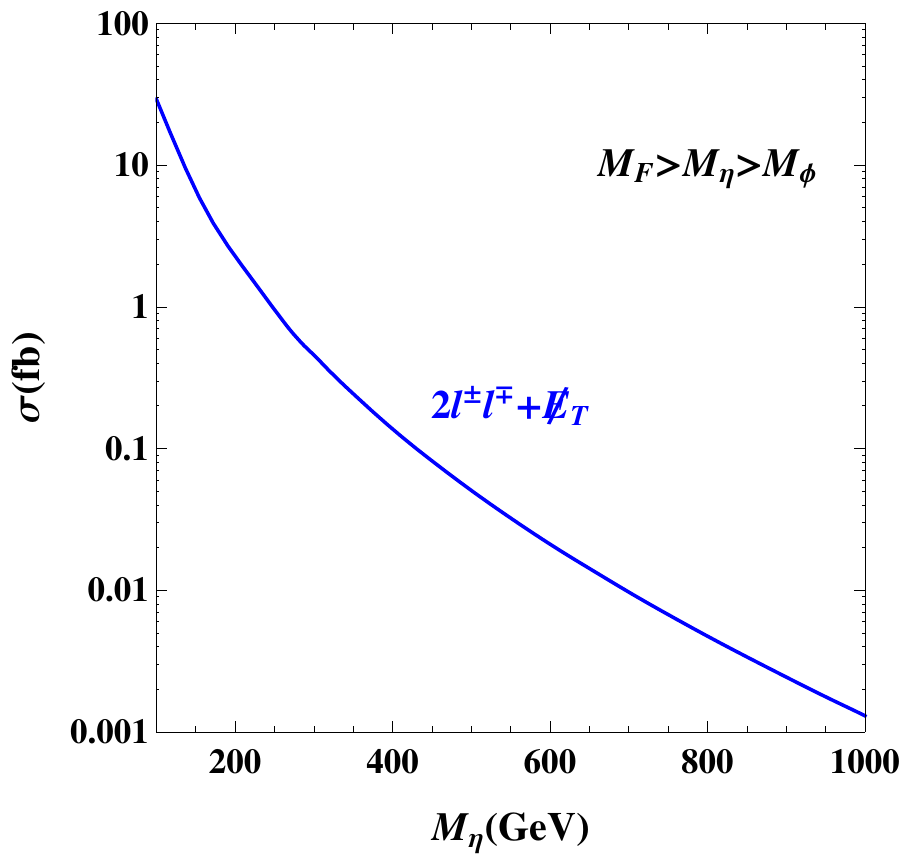}
\end{center}
\caption{Cross section of dilepton signature $\ell^+\ell^-+\cancel{E}_T$ (left) and trilepton signature $2\ell^\pm\ell^\mp+\cancel{E}_T$ (right) at $14~\TeV$ LHC.
\label{Fig:MAS}}
\end{figure}

As for model (a), since both $F\sim(1,1,0)$ and $\phi\sim(1,1,0)$
are pure singlet, only the inert doublet
$\eta=(\eta^+,(\eta^0_R+\eta^0_I)/\sqrt{2})^T\sim(1,2,1)$ can be
pair produced at LHC
\begin{equation}
pp\to \eta^+\eta^-, \eta^0_R\eta^0_I, \eta^\pm \eta^0_{R,I}.
\end{equation}
Because there are always a pair of DM in the final states, the
signatures will thus contains missing transverse energy
$\cancel{E}_T$. Here, we consider multi-lepton plus $\cancel{E}_T$
signatures. For inert doublet DM $\eta^0_R/\eta^0_I$, multi-lepton
plus $\cancel{E}_T$ signatures at LHC have been extensively studied
in Ref. \cite{Dolle:2009ft}, thus we concentrate on $F$ or $\phi$
DM. For fermion singlet DM, the promising signature is
\begin{equation}
pp\to \eta^+\eta^- \to \ell^+ F + \ell^- F,
\end{equation}
which leads to $\ell^+\ell^-+\cancel{E}_T$ signature at LHC.
Cross section of this dilepton signature is presented in left
panel of Fig.~\ref{Fig:MAS}.
Searches for such dilepton signature has been performed by ATLAS
\cite{Aad:2014vma} and CMS \cite{Khachatryan:2014qwa}. Assuming
$\eta^\pm$ exclusive decays into $e^\pm F$ or $\mu^\pm F$, ATLAS has
excluded the region with $M_{\eta^\pm}\lesssim300~\GeV$ and
$M_{F}\lesssim150~\GeV$ \cite{Aad:2014vma}, meanwhile the CMS limit
is less stringent \cite{Khachatryan:2014qwa}. On the other hand, for
scalar singlet DM, the promising signature is
\begin{equation}
pp\to \eta^\pm \eta^0_{R,I} \to W^\pm \phi + Z \phi \to 2\ell^\pm
\ell^\mp + \cancel{E}_T.
\end{equation}
Cross section of this dilepton signature is presented in right
panel of Fig.~\ref{Fig:MAS}.
Searches for such trilepton signature has also been performed by
ATLAS \cite{Aad:2014nua} and CMS \cite{Khachatryan:2014qwa}. The
more stringent limit is also set by ATLAS, with
$M_{\eta^\pm}\lesssim350~\GeV$ and $M_{\phi}\lesssim120~\GeV$ being
excluded \cite{Aad:2014nua}. Note that this exclusion limit is
acquired in simplified SUSY model with chargino-neutralino
associated production. The exclusion limit is expected weaker in
model (a), mainly because the cross section of
$\eta^\pm\eta^0_{R,I}$ is much smaller than the cross section of
chargino-neutralino with same masses.

Before ending this section, we give one benchmark point for each of
the two benchmark models by considering the constraints from
the phenomenologies we just discussed above. First, for the tree level
benchmark model (B), the benchmark point is
\begin{equation}
y_1=y_2=0.01,~M_F=M_\phi=1~\TeV, \langle \phi \rangle=10~\keV.
\end{equation}
Then, for the one-loop level benchmark model (a), the benchmark point is
\begin{eqnarray}
y_1^{i1}=y_2^{i1}=10^{-6},~y_1^{i2,i3}=y_2^{i2,i3}=10^{-2},\theta=0.01,\\\nonumber
M_\phi=60~\GeV,~M_\eta=200~\GeV,~M_F=10^7~\GeV.
\end{eqnarray}

\section{Conclusion}\label{CL}
With at most two additional scalars and a heavy intermediate
fermion, we perform a systematical study on pathways that can
naturally generate tiny Dirac neutrino masses at tree- and one-loop
level, In both cases, we concentrate on the $SU(2)_L$ scalar
multiplet no larger than quintuplet, and derive the complete sets of
viable models.

To realize tree level models in Fig.~\ref{FM:Tree}, the conservation
of lepton number symmetry is assumed to forbid the unwanted Majorana
mass term $(m_N/2)\overline{\nu_R^C}\nu_R$. Then an extra $Z_2$
symmetry is employed to forbid direct
$\bar{\nu}_L\nu_R\overline{\phi^0}$ coupling. The breaking of
this $Z_2$ symmetry will induce an effective
small Dirac neutrino mass term $m_D\bar{\nu}_L\nu_R$. For tree level
models, a finite set of model is found by requiring the natural
small VEVs of new scalars. If one of the added scalars is actually
the SM Higgs fields $H$ itself, there are four types of
realizations, which correspond to $d=6$ effective low-energy
operators. On the other hand, if two new scalars are introduced,
then they are usually triplets/quadruplets/qintuplets for minimal
models, corresponding to $d=8$ or $d=10$ effective low-energy
operators.

To realize purely radiative models in Fig.~\ref{FM:Loop}, we further
impose $Z_2^D$ symmetry, under which $S_{1,2}$ and $F$ carry
$Z_2^D$-odd charge while all SM fields transform trivially. The
lightest particle within the inert fields $S_{1,2}$ and $F$ is
stable, and thus becomes a dark matter candidate if it has no
electric charge. We exhaust the list of viable models, and briefly
discuss the possible DM candidate. Note that current direct
detection limits have already excluded some models. Clearly, for
fermion DM, it could be $F\sim(1,1,0),(1,3,0),(1,5,0)$, while for
scalar DM, we have more option, e.g., the inert doublet
$\eta\sim(1,2,1)$. The important fact is that if the DM candidate
has non-zero hyper charge, a mixing between $S_1$ and $S_2$ and/or a
quartic term as $(S_{1,2}H)^2$ is required to induce a large enough
mass splitting between the real and imaginary part of the neutral
component.

As for the phenomenological issues, the Yukawa coupling $y_1
\overline{F_R} L_L S_1$ will induce lepton flavor violation (LFV)
processes. For tree level models, current limits on BR($\mu \to
e\gamma$) denotes that $M_{S_1} \langle S_1 \rangle \gtrsim
600~\GeV\cdot\MeV$, with $y_1=y_2$ and $\langle S_1 \rangle
=\langle S_2 \rangle$ being assumed. Meanwhile for
radiative models, tight constraints from LFV requires the Yukawa
coupling $|y_1|\lesssim0.01$ when both $F$ and $S_1$ are located
around electroweak scale. On the other hand, if $F$ is heavy enough,
i.e., $M_F\sim10^7~\GeV$, which is also possible in radiative models,
the leptogenesis is also possible. For the scalar singlet DM $\phi$ in
model (a), we perform a brief discussion on relic density,
direct detection, indirect detection and Higgs invisible decay.
And we find the the only allowed region is $M_\phi\lesssim M_h/2$.
 To illustrate LHC signatures, we
take model (B) and model (a) as an example. For tree level model
(B), the promising signatures is trilepton signature $\Sigma^\pm \to
\ell^\pm Z \to \ell^\pm \ell^+\ell^-$ when $M_{\Sigma}<M_{\phi}$.
While for $M_{\Sigma}>M_{\phi}$, the five-lepton signature
$\Sigma^\pm \to \ell^\pm \phi \to \ell^\pm ZZ\to3\ell^\pm2\ell^\mp$
might be promising. In case of loop level model (a), the promising
signature is $\ell^+\ell^-+\cancel{E}_T$ if $F$ is DM, and
$2\ell^\pm\ell^\mp+\cancel{E}_T$ if $\phi$ is DM.

\section*{Acknowledgments}
The work of Weijian Wang is supported by National Natural Science
Foundation of China under Grant Numbers 11505062, Special Fund of
Theoretical Physics under Grant Numbers 11447117 and Fundamental
Research Funds for the Central Universities. The work of Zhi-Long Han
is supported in part by the Grants No. NSFC-11575089.

\newpage
\section*{Appendix A: One-loop Neutrino Mass models with Larger Multiplets}

\begin{table}[!htbp]
\begin{center}
\begin{tabular}{|c|c|c|c|c|}\hline\hline
Models & $F$ & $S_{1}$ & $S_{2}$ & $Z_2^D$ DM\\
\hline (k) & $(1,3,0)$ & $(1,4,1)$ & $(1,3,0)$ & Inert Triplet or Quadruplet\\
\hline (l) & $(1,3,-2)$ & $(1,4,-1)$ & $(1,3,2)$ & Inert Triplet or Quadruplet\\
\hline (m) & $(1,3,2)$ & $(1,4,3)$ & $(1,3,-2)$ & Excluded\\
\hline (n) & $(1,3,-4)$ & $(1,4,-3)$ & $(1,3,4)$ & Excluded\\
\hline (o) & $(1,4,-1)$ & $(1,4\pm1,0)$ & $(1,4,1)$ & \makecell{Inert Triplet/Quintuplet\\ or Quadruplet}\\
\hline (p) & $(1,4,1)$ & $(1,4\pm1,2)$ & $(1,4,-1)$ & \makecell{Inert Triplet/Quintuplet\\ or Quadruplet}\\
\hline (q) & $(1,4,-3)$ & $(1,4\pm1,-2)$ & $(1,4,3)$ & Excluded\\
\hline (r) & $(1,4,3)$ & $(1,4\pm1,4)$ & $(1,4,-3)$ & Excluded\\
\hline (s) & $(1,4,-5)$ & $(1,5,-4)$ & $(1,4,5)$ & Excluded\\
\hline (t) & $(1,5,0)$ & $(1,4,1)$ & $(1,5,0)$ & Inert Quadruplet or Quintuplet\\
\hline (u) & $(1,5,-2)$ & $(1,4,-1)$ & $(1,5,2)$ & Inert Quadruplet or Quintuplet\\
\hline (v) & $(1,5,2)$ & $(1,4,3)$ & $(1,5,-2)$ & Excluded\\
\hline (w) & $(1,5,-4)$ & $(1,4,-3)$ & $(1,5,4)$ & Excluded\\
\hline (x) & $(1,5,4)$ & $(1,4,5)$ & $(1,5,-4)$ & Excluded\\
\hline\hline
\end{tabular}
\caption{Radiative neutrino mass for Dirac neutrinos with quadruplet or/and quintuplet and DM candidate.} \label{tabnew3}
\end{center}
\end{table}

\section*{Appendix B: Dark Matter Annihilation Cross Sections}

Annihilation into SM fermions:
\begin{equation}
\sigma(\phi\phi\to f\bar{f})v_{\text{rel}}=
\frac{\lambda_\phi^2 M_f^2 N_c^f (1-4M_f^2/s)^{3/2}}
{\pi\left[(s-M_h^2)^2+M_h^2\Gamma_h^2\right]},
\end{equation}
where $N_c^f$ is the color factor for fermion $f$.

Annihilation into $W^+W^-$:
\begin{equation}
\sigma(\phi\phi\to W^+W^-) v_{\text{rel}}=
\frac{\lambda_\phi^2(s^2-4M_W^2 s + 12 M_W^4)\sqrt{1-4M_W^2/s}}
{2\pi s \left[(s-M_h^2)^2+M_h^2\Gamma_h^2\right]}.
\end{equation}

Annihilation into $ZZ$:
\begin{equation}
\sigma(\phi\phi\to ZZ) v_{\text{rel}}=
\frac{\lambda_\phi^2(s^2-4M_Z^2 s + 12 M_Z^4)\sqrt{1-4M_Z^2/s}}
{4\pi s \left[(s-M_h^2)^2+M_h^2\Gamma_h^2\right]}.
\end{equation}

Annihilation into $hh$ in the $s\to 4M_\phi^2$ limit:
\begin{equation}
\sigma(\phi\phi\to hh)v_{\text{rel}}=
\frac{\lambda_\phi^2 \left[M_h^4-4M_\phi^2+2\lambda_\phi v^2 (4M_\phi^2-M_h^2)\right]^2}
{4\pi M_\phi^2 \left(M_h^4 - 6M_h^2 M_\phi^2 + 8 M_\phi^2\right)^2}\sqrt{1-\frac{M_h^2}{M_\phi^2}}.
\end{equation}

Annihilation into $\gamma\gamma$:
\begin{equation}
\sigma(\phi\phi\to \gamma\gamma) v_{\text{rel}}=
\frac{16 \lambda_\phi^2 v^2 \Gamma_{\gamma\gamma}(s)}
{\sqrt{s} \left[(s-M_h^2)^2+M_h^2\Gamma_h^2\right]},
\end{equation}
where the width $\Gamma_{\gamma\gamma}(s)$ is given by:
\begin{equation}
\Gamma_{\gamma\gamma}(s)=\frac{\alpha^2 s^{3/2}}{512\pi^3 v^2 }
\left|\sum_f N_c^f Q_f^2 A_{1/2}(\tau_f)+A_1(\tau_W)\right|^2,
\end{equation}
with $\tau_i=s/(4M_i^2)$ and the form factor:
\begin{eqnarray}
A_{1/2}(\tau) & = & 2 [\tau + (\tau-1) f(\tau)]\tau^{-2},\\
A_{1}(\tau) & = &  -[2\tau^2+3\tau 3(2\tau-1)f(\tau)]\tau^{-2},
\end{eqnarray}
where $f(\tau)$ is
\begin{eqnarray}
f(\tau)=\left\{
\begin{array}{ll}
\arcsin^2\sqrt{\tau} &\quad\tau\leq1\\
-\frac{1}{4}\left[\log\frac{1+\sqrt{1-\tau^{-1}}}{1-\sqrt{1-\tau^{-1}}}-i\pi\right]^2 &\quad\tau>1
\end{array} \right.
\end{eqnarray}

\end{document}